\documentclass{article}
\pdfoutput=1
\usepackage{graphicx}	% Including figure files
\usepackage{amsmath}	% Advanced maths commands
\usepackage{amssymb}	% Extra maths symbols
\usepackage{wasysym}    % Moon symbol
\usepackage[round]{natbib}
\usepackage{authblk}
\usepackage{caption}

\usepackage[utf8]{inputenc}
\usepackage{combelow}% provides \cb to place comma below character
\usepackage{newunicodechar}

\newunicodechar{ș}{\cb{s}}

\author[1]{Fabien Gachet}
\author[1]{Alessandra Celletti}
\author[1]{Giuseppe Pucacco}
\author[2]{Christos Efthymiopoulos}

\affil[1]{\emph{Department of Mathematics, University of Rome Tor Vergata, Rome, Italy}}
\affil[2]{\emph{Research Center for Astronomy and Applied Mathematics, Academy of Athens, Athens, Greece}}

\title{Geostationary secular dynamics revisited:\\ application to high area-to-mass ratio objects}

\date{}

\begin{document}

\maketitle

\begin{abstract}
The long-term dynamics of the geostationary Earth orbits (GEO) is revisited through the application of canonical perturbation theory.
We consider a Hamiltonian model accounting for all major perturbations: geopotential at order and degree two, lunisolar perturbations with a realistic model for the Sun and Moon orbits, and solar radiation pressure.
The long-term dynamics of the GEO region has been studied both numerically and analytically, in view of the relevance of such studies to the issue of space debris or to the disposal of GEO satellites.
Past studies focused on the orbital evolution of objects around a nominal solution, hereafter called the forced equilibrium solution, which shows a particularly strong dependence on the area-to-mass ratio.
Here, we i) give theoretical estimates for the long-term behavior of such orbits, and ii) we examine the nature of the forced equilibrium itself.
In the lowest approximation, the forced equilibrium implies motion with a constant non-zero average `forced eccentricity', as well as a constant non-zero average inclination, otherwise known in satellite dynamics as the inclination of the invariant `Laplace plane'.
Using a higher order normal form, we demonstrate that this equilibrium actually represents not a point in phase space, but a trajectory taking place on a lower-dimensional torus.
We give analytical expressions for this special trajectory, and we compare our results to those found by numerical orbit propagation.
We finally discuss the use of proper elements, i.e., approximate integrals of motion for the GEO orbits.
\end{abstract}

%%%%%%%%%%%%%%%%%%%%%%%%%%%%%%%%%%%%%%%%%%%%%%%%%%%%%%%%%%%%%%%%%%%%%%%%%%%%%%%
\section{Introduction}
\label{sec:intro}
%%%%%%%%%%%%%%%%%%%%%%%%%%%%%%%%%%%%%%%%%%%%%%%%%%%%%%%%%%%%%%%%%%%%%%%%%%%%%%%
The secular (long-term) dynamics of natural or artificial satellites of planets is governed by interactions including the multipole gravitational attraction of the main planet, the attraction of distant third bodies and non gravitational forces, e.g. the solar radiation pressure, drag forces etc.
Apart from the main planet's monopole potential term, all remaining forces can be practically considered as small perturbations. Thus, besides purely numerical approaches, the study of secular dynamics can be reduced to an analytical problem of perturbation theory including many small parameters.\\
In the present paper, we introduce a Hamiltonian approach for the analytical study of satellite secular dynamics, using a technique based on  {\it Hamiltonian normal forms}.
To this end, we focus on the Geostationary Earth Orbit (GEO) resonant domain, where dissipative forces are nearly negligible and all forces can be modeled to a good first approximation using a Hamiltonian model.
Our particular motivation for this study is the need to model the long-term uncontrolled evolution of {\it space debris}.
As will be discussed below, for particular populations of space debris called high area-to-mass ratio (HAMR) objects, this evolution depends crucially on the solar radiation pressure term. Yet, our choice of variables and normal form technique used below apply to both low and high area-to-mass ratio objects.
Thus, more general satellite dynamics problems of conservative character (see for example \citet{MignardHenon1984} or \citet{Tremaine2009}), can be treated by the same technique.

Returning to the GEO problem, as shown by numerical studies of the long-term evolution of typical satellites with a low $A/m$ close to GEO \citep{Anselmo2008}, the dynamics in this region is governed by the action of the low-order terms of the Earth's gravitational field and the lunisolar gravitational perturbations \citep{Chao2005,Celletti2014}.
Unraveling the long-term dynamics is more complicated in the case of space debris. In particular, a new class of high area-to-mass ratio (HAMR) debris, with quite peculiar dynamics compared to more typical low $A/m$ objects, was discovered a decade ago \citep{Schildknecht2004}.
Depending on their generation process (see \citet{Liou2005}), such objects can reach apparent $A/m$ ratios as high as 30 m\textsuperscript{2}kg\textsuperscript{-1} \citep{Schildknecht2007}.
As shown in \citet{Anselmo2005}, the HAMR objects can oscillate from zero to a value of the eccentricity as high as $\sim 0.5$ with a period of about one year, an effect mainly due to the action of the solar radiation pressure (SRP); their inclination can also reach values as high as 30 to 40 degrees, while the associated period (precession of the nodes) can decrease to less than 20 years.
These phenomena have been studied analytically by \citet{ChaoBaker,Chao2006,Valk2008a,Rosengren2013,Casanova2014}.
These studies show that at GEO, SRP modifies considerably the equilibrium of the dynamical system.
Concerning the inclination, the Laplace plane, around which the libration takes place, undergoes an increase in its inclination from 7.3 deg for objects with low $A/m$ \citep{AllanCook1964}, to higher values directly depending on $A/m$, as noticed in \citet{AllanCook1967} and explicited in \citet{Rosengren2014b}, where implicit equations are given.
As for the eccentricity, SRP causes large yearly variations around a mean value, i.e., the \emph{forced eccentricity}.
The latter can be elegantly approximated by $e_{forced} \sim 0.01 \times C_r  \frac{A}{m}$, where $C_r$ is the reflectivity coefficient \citep{Valk2008a}
%------------------------------------------
\footnote{\label{foot:Cr}
One has $1\leq C_r\leq 2$, $C_r=1$ for a perfect absorber, and $C_r=2$ for a perfect mirror.
In practice, only the product $C_r \frac{A}{m}$ is needed in order to determine the SRP force. As in \citet{Valk2008a}, in this study we set $C_r=1$, so that all results are parameterized in terms of an effective $A/m$, equal in reality to the product $C_r \frac{A}{m}$.}.
%------------------------------------------
For debris generated from a parent body originally in a nearly circular orbit, the yearly variation of the eccentricity is about twice the value of $e_{forced}$, i.e., again, proportional to $A/m$.
Actually, the variations of the eccentricity and the inclination due to SRP are coupled as shown in \citet{Valk2008a}, and the long-term change in inclination adds another period in the movement of the eccentricity altogether, equal to the precession period of the longitude of the ascending node.
Some of these results were obtained numerically already in \citet{Friesen1992}.

In the present paper, we revisit the problem of the long-term dynamics in the GEO region, aiming to provide a new method by which the nature of the orbits can be unraveled using analytical approaches, most notably the method of Hamiltonian normal forms.
Besides this basic novel formulation, our study presents two more differences with respect to previous analytical approaches referenced above.
i) We use a natural system of coordinates (Earth-fixed, Earth-centered cylindrical coordinates), which simplify considerably all the expansions, compared to expansions in terms of orbital elements.
ii) We consider realistic lunisolar orbits (e.g., with four independent frequencies for the Moon).
Regarding (i), although it has become a tradition in Celestial Mechanics to use the language of elements for analytical studies of perturbed Kepler problems, the fact should be stressed that, for orbits not too highly eccentric or highly inclined, the use of post-epicyclic approximations performs equally well, while simplifying considerably the algebra.
This last point is particularly appreciated when high-order expansions are to be made, as is the case with our current technique demonstrated in Section \ref{sec:hamexp} below.
Regarding (ii), our final model, exposed in Section \ref{sec:model}, has eight degrees of freedom, while our normal form algorithm is designed so as to represent analytically the space-time propagation of objects moving in the GEO domain.
In order to facilitate a comparison with other propagation routines, we provide a full form of our Hamiltonian model in electronic format as a supplementary material to this paper.

As an example of the new approach, we derive analytical expressions for a very basic orbit in the GEO domain, i.e., the forced equilibrium GEO solution.
Using canonical perturbation theory, we perform an averaging of the 8 degrees of freedom Hamiltonian by the method of normal forms via Lie series \citep{Hori1966,Deprit1969}.
The forced equilibrium is represented by a stable equilibrium point of the final averaged Hamiltonian, i.e., the normal form. The position of this equilibrium in phase space allows to recover analytical results found in  \citet{ChaoBaker,Chao2006,Valk2008a,Rosengren2013}, and even to give more precise formulas since the normal form obtained is of higher order.
However, the main new information regarding the nature of the forced equilibrium solution comes from back-transforming the normal form equilibrium solution to the one in the original variables.
Due to this transformation, we find that the forced equilibrium solution will appear in real space as a quasi-periodic trajectory with a spectrum of frequencies (five basic ones, in our model, and their multiples).
Furthermore, the amplitude of the oscillation depends on the size of all perturbations, most notably it depends sensitively on the area-to-mass ratio $A/m$.
An analytical expression of this basic trajectory is derived as a function of time.
This allows to obtain also initial conditions lying on a lower-dimensional (5-dimensional) invariant torus embedded in the 8-dimensional phase space, whose otherwise identification by purely numerical means represents a formidable task.

As a demonstration we compare our analytical trajectories of the forced equilibrium solution with a numerical computation of the trajectories up to 100 years for different $A/m$ ratios.
We find an error of $\sim 1\%$ in times of the order of the secular period (tens of years), mainly due to the relatively low-order truncation of our given normal form example (an analysis of the time evolution of the error in terms of the truncation order is done in Section \ref{sec:epi}).
However, we also point out the effect on analytical computation due to the presence of particular \emph{small divisors} representing near-resonances between the various frequencies entering the normal form calculations.

Overall, we would like to emphasize the practical aspects and benefits of calculations based on a detailed normalization process as the one proposed below. The main benefit comes from acquiring an explicit representation of the canonical mapping between the original (cartesian) phase space variables 
of the test particle and those found after the normalization (e.g. via Lie series). The reverse mapping allows, in turn, to express the dynamics in the original variables, thus rendering the whole approach amenable to direct comparison with numerical propagators of the orbits. As discussed below, the analytical approximation based on normal forms, albeit less accurate than a numerical one, yields quite precise initial conditions for particular classes of orbits (as, for example, `safe disposal' orbits for space debris). The latter could then be further refined by numerical means. On the other hand, the perturbative approach provides insight to the long term dynamical behavior of space debris in a way which appears hardly attainable by purely numerical means.

The paper is structured as follows: Section~\ref{sec:model} presents our Hamiltonian model, the coordinate system employed, as well as the adopted forms of the lunisolar and SRP perturbations.
Section~\ref{sec:hamexp} presents the application of the canonical perturbation theory, construction of the normal form and explicit formulas for the long-term dynamics around the equilibrium solution.
Section~\ref{sec:numres} presents the application of the analytical computation of the low-dimensional torus solution of the forced equilibrium and its comparison with numerical results.
Finally, Section~\ref{sec:concl} summarizes the main conclusions of the present study.

%%%%%%%%%%%%%%%%%%%%%%%%%%%%%%%%%%%%%%%%%%%%%%%%%%%%%%%%%%%%%%%%%%%
\section{Model}\label{sec:model}
%%%%%%%%%%%%%%%%%%%%%%%%%%%%%%%%%%%%%%%%%%%%%%%%%%%%%%%%%%%%%%%%%%%
We consider a body orbiting the Earth in the geostationary region and subject to the action of the following forces, which are given to some approximation as described below:
\begin{itemize}
\item[(i)] the gravitation of the Earth, including the oblateness ($J_2$, or equivalently, $C_{2,0}$ term) and the equatorial ellipticity of the Earth (tesseral harmonics $C_{2,2}$ and $S_{2,2}$);
\item[(ii)] the gravitational perturbations due to the Moon and the Sun, whose potentials are expanded up to order 2 in the ratio of the geocentric distance of the satellite and of the Sun and to order 4 in the ratio of the geocentric distance of the satellite and of the Moon;
\item[(iii)] the solar radiation pressure (SRP) perturbations, using the \emph{cannonball} approximation \citep{Kubo1999}. We do not consider the effect of the Earth's shadow on the satellite; as shown in \citet{Hubaux2013} this effect does not contribute significantly to the dynamics. Models more detailed than the cannonball approximation have been proposed in the literature (e.g. \citet{McMahon2010}) but their use in the Hamiltonian context is less practical. On the other hand, the cannonball model cannot capture the effect on SRP of the  orientation of the body. However, regarding long-term averages, the cannonball model appears sufficient in the analytical study of secular dynamics \citep{Rosengren2013,Rosengren2014b}.
\end{itemize}
It is important to note that, at GEO, different perturbations end up having a comparable  magnitude, as can be seen in \citet{Valk2008a}. Also, restricting the study of the SRP effect only in the direction along the line joining the test particle with the Sun, the SRP can be shown to effectively acquire the form of a conservative force. Exploiting this fact, in the following subsections we first define our coordinate system, and then express all the above forces in the context of the canonical formalism, i.e., using a set of elementary canonical transformations and deriving the form of the Hamiltonian function which generates the equations of motion. 

%------------------------------------------------------
\subsection{Hamiltonian}\label{sec:HamDef}
%------------------------------------------------------

We consider a Cartesian inertial frame of reference $(x,y,z)$ whose origin is at the center of the Earth at the Universal Time 12 hours (noon) of the Julian day 1 JAN 2000, considered in the following as the origin of time, i.e., as $t=0$. The $z$-axis coincides with the Earth's polar axis, while the $x-$axis points toward the position of the Greenwich Meridian at this epoch.   
The above inertial frame differs from the familiar Earth Mean Equator and Equinox of J2000 (EME2000) geocentric inertial frame (see \citet{Montenbruck2000}, p.170) by a simple rotation with respect to the $z$-axis, yielding the angle between the direction of the mean equinox and that of Greenwich at our specific epoch. Thus, if $(X,Y,Z)$ are EME2000 Cartesian 
coordinates, we have $(x,y,z)^T=R_{G} (X,Y,Z)^T$ where
\begin{equation}\label{rg}
R_{G}=\left(
\begin{array}{ccc}
\cos \Omega_{G} & \sin \Omega_{G} & 0 \\
-\sin\Omega_{G} & \cos \Omega_{G} & 0\\
0 & 0 & 1
\end{array} \right)\ ,
\end{equation}
with $\Omega_{G}=280.4606^\circ$ \footnote{The parameters used in our study are presented with as many digits as given in the related literature references. However, in the final Hamiltonian expansion, given as a supplementary material (see below), all numerical coefficients are truncated at the level of $10^{-10}$.}.

We define cylindrical coordinates $(\rho,\Phi,z)$ in the above frame via $x=\rho\cos\Phi$, $y=\rho\sin\Phi$. As detailed below, parameterizing SRP in terms of the $A/m$ ratio allows to eliminate the mass of the test particle from the equations of motion. Then, the latter are given by Hamilton's equations under the Hamiltonian
\begin{equation}\label{eq:HamSphPre}
H\equiv H(\rho,\Phi,z,p_\rho,p_\Phi,p_z,t)=
\frac{{p_\rho}^2}{2}+\frac{{p_\Phi}^2}{2\rho^2}+\frac{p_z^2}{2}+V(\rho,\Phi,z,t)\ ,
\end{equation}
where
\begin{eqnarray}\label{eq:momenta}
p_\rho=\dot{\rho},~~~p_\Phi=\rho^2\dot{\Phi},~~~p_z=\dot{z}~,  \nonumber
\end{eqnarray}
and $V$ represents the potential derived from all forces accounted for in the model.

We now analyze the form of the function $V(\rho,\Phi,z,t)$, while the dependence on time of the Hamiltonian is dealt with by an appropriate set of canonical transformations as explained in subsection \ref{sec:Hdeptime}.

The function $V$ is decomposed as:
\begin{equation}\label{pot}
V=V_{GEO_2}+V_{\leftmoon}+V_\odot+V_{SRP}
\end{equation}
where $V_{GEO_2}$ is the geopotential, $V_{\leftmoon}$, $V_\odot$ the gravitational perturbation potentials of the Moon and Sun respectively, and $V_{SRP}$ the solar radiation pressure potential. These terms are given in detail as follows.

%---------------------------------------
\subsubsection{Geopotential}
%---------------------------------------
We consider the expansion of the geopotential in spherical harmonics \citep{Kaula1966} up to quadrupole terms. Let $\varphi=\Phi-\Omega_E t$, where 
\begin{equation}\label{omeearth}
\Omega_E = 7.292115\times 10^{-5}\mbox{rad}~\mbox{s}^{-1} = 131850.9^\circ \ \text{yr}^{-1}
\end{equation}
is the sidereal rotation frequency of the Earth \citep{EGM96}. Then
\begin{equation}\label{eq:VGEO2}
\begin{aligned}
V_{GEO_2}&=-\frac{\mu _\oplus}{\sqrt{\rho^2+z^2}}+\frac{\sqrt{5} \bar C_{2,0} \mu _\oplus R_\oplus^2}{2(\rho^2+z^2)^{3/2}}-\frac{3 \sqrt{5} \bar C_{2,0} \mu _\oplus R_\oplus^2 z^2}{2(\rho^2+z^2)^{5/2}}\\
&-\frac{\sqrt{15} \mu _\oplus R_\oplus^2}{2(\rho^2+z^2)^{3/2}}\left(1-\frac{z^2}{\rho^2+z^2}\right)\left(\bar C_{2,2}  \cos (2 \varphi ) +\bar S_{2,2} \sin (2 \varphi) \right)
\end{aligned}
\end{equation}
where $\mu_\oplus=398600.4418 \, \text{km}^3 \text{s}^{-2}$ and $R_\oplus=6378.137 \,\text{km}$ are the standard gravitational parameter and the equatorial radius of the Earth \citep{EGM96}, and $\bar C_{n,m}$, $\bar S_{n,m}$ the normalized spherical harmonic coefficients. The physical values of the coefficients are given in Table \ref{tab:coeffgeopot}. Note that the coefficients of degree 1 are 0 since the center of mass is at the origin of the coordinate system, and the coefficients of degree 2 and order 1 are 0 since the $z$ axis is aligned with the rotation of the Earth in our model.
\begin{table}
\centering
\caption{Values of $\bar C_{n,m}, \bar S_{n,m}$ in units of $10^{-6}$ up to degree and order 2, from NASA EGM96 \citep{EGM96}.}\vspace{2mm}
\begin{tabular}{llll}
\hline\hline
\rule{0pt}{3ex} $n$& $m$& $\bar C_{n,m}\times 10^6$& $\bar S_{n,m}\times 10^6$\\
\hline
2& 0& -484.165371736& 0\\
2& 2& 2.43914352398& -1.40016683654\\
\hline\hline
\end{tabular}
\label{tab:coeffgeopot}
\end{table}

%-----------------------------------------------------
\subsubsection{Lunisolar gravitational perturbations}
\label{sec:lunisolar}
%-----------------------------------------------------
The lunisolar perturbations are described by the following potential terms:
\begin{equation}
\label{eq:potsun}
V_\odot=-\mu_\odot \left( \frac{1}{|\mathbf{r}-\mathbf{r_\odot}|} - \frac{\mathbf{r}\cdot\mathbf{r_\odot}}{|\mathbf{r_\odot}|^3} \right)\ ,
\end{equation}
and
\begin{equation}
\label{eq:potmoon}
V_{\leftmoon}=-\mu{\leftmoon} \left( \frac{1}{|\mathbf{r}-\mathbf{r_{\leftmoon}}|} - \frac{\mathbf{r}\cdot\mathbf{r_{\leftmoon}}}{|\mathbf{r_{\leftmoon}}|^3} \right)\ .
\end{equation}
In these expressions we have:

i) the constants $\mu_\odot= 1.32712440018 \times 10^{11} \, \text{km}^3 \text{s}^{-2}$ and $\mu_{\leftmoon}= 4902.8000 \, \text{km}^3 \text{s}^{-2}$  (standard gravitational parameter of the Sun and the Moon respectively \citep{Roncoli2005}).

ii) The test particle's position vector
\begin{equation}\label{rvec}
\mathbf{r}=\left(
\begin{array}{c}
\rho \cos \Phi \\
\rho \sin \Phi\\
z
\end{array} \right)
=\left(
\begin{array}{c}
\rho \cos (\varphi+\varphi_E) \\
\rho \sin (\varphi+\varphi_E)\\
z
\end{array} \right)
\end{equation}
with
\begin{equation}
\varphi_E=\Omega_E t \ .
\end{equation}

iii) The Sun and Moon time-varying position vectors $\mathbf{r_\odot}$, $\mathbf{r_{\leftmoon}}$. To obtain these functions of time, we use the expressions given in \citet{Montenbruck2000} (p.71-72), recalled in Appendix A, with formulas truncated from series expansions up to second order in the eccentricities and inclinations, but still accurate to 0.1-1\% for decades around the year 2000.

The combination of angles $\varphi_M,\varphi_{M_a},\varphi_{M_p},\varphi_{M_s}$ and their associated rates \eqref{eq:ratesun}, \eqref{eq:rates} have been defined so that they correspond to the most important features of these orbits, known since Babylonian astronomy, namely: the yearly frequency with which the Sun revolves around the Earth in the geocentric frame $\Omega_M$, the orbital revolution of the Moon around the Earth in a Lunar month (here we took the anomalistic month of 27.55 days as a reference) $\Omega_{M_a}$, the precession of the Moon's argument of periapse with a period of about 8.85 years $\Omega_{M_p}$ and the precession of the Moon's ascending node with a period of about 18.6 years $\Omega_{M_s}$.

%-----------------------------------------------------
\subsubsection{The solar radiation pressure}
\label{sec:SRP}
%-----------------------------------------------------
For long-term dynamics, the model typically used to describe the solar radiation pressure perturbation is the cannonball model \citep{Rosengren2013}. The object is treated as a sphere with constant reflectivity and $A/m$ ratio, the total momentum imparted by the solar photons is assumed to be independent of the object's attitude, and the resultant force is along the Sun-object line. The solar radiation pressure perturbation is then derived from the following potential:
\begin{equation}
\label{eq:SRPfull}
V_{SRP}=C_rP_r\ a_\odot^2\ \frac{A}{m}\ \frac{1}{|\mathbf{r}-\mathbf{r_\odot}|}
\end{equation}
with $C_r$ the reflectivity coefficient (we set $C_r=1$, see footnote \ref{foot:Cr}), $ P_r=4.56\times 10^{-6}$~Nm\textsuperscript{-2} the radiation pressure for an object located at $a_\odot=1$AU, the astronomical unit of distance, and $A/m$ the area-to-mass ratio.
We point out that the model we consider for the Sun, moving on an inclined ellipse, is crucial for SRP since, as noted in \citet{Valk2008a}, having a fixed Sun-Earth distance in the estimation of SRP (an assumption made in some of the previous studies such as \citet{Chao2006}) would induce spurious long-period terms in eccentricity and inclination evolution.

%-------------------------------------------------------------------------------
\subsection{Removal of the explicit time dependence}
\label{sec:Hdeptime}
%-------------------------------------------------------------------------------
The Hamiltonian \eqref{eq:HamSphPre} depends on the particle's three coordinates $\rho,\Phi,z$ and conjugate momenta $p_\rho$, $p_\Phi$, $p_z$, and also it depends explicitly on time through trigonometric terms depending on quantities $\varphi_E=\Omega_E t$, $\varphi_M = \Omega_M t$, $\varphi_{M_a} = \Omega_{M_a} t$, $\varphi_{M_p}=\Omega_{M_p} t$, and $\varphi_{M_s}=\Omega_{M_s} t$. Following a standard process, we can augment the number of degrees of freedom in the Hamiltonian by promoting each of the angles $(\varphi_E,\varphi_M,\varphi_{M_a},\varphi_{M_p},\varphi_{M_s})$ as a new canonical position variable, associated with a conjugate momentum variable, i.e., the so-called `dummy action' variables $(I_E,J_M,J_{M_a},J_{M_p},J_{M_s})$. The latter variables appear in the `extended Hamiltonian'
\begin{equation}\label{eq:HamSphExt}
\begin{aligned}
&H\equiv H(\rho,\Phi,z,\varphi_E,\varphi_M,\varphi_{M_a},\varphi_{M_p},\varphi_{M_s},p_\rho,p_\Phi,p_z,I_E,J_M,J_{M_a},J_{M_p},J_{M_s})\\
&=\frac{{p_\rho}^2}{2}+\frac{{p_\Phi}^2}{2\rho^2}+\frac{p_z^2}{2}
+V(\rho,\Phi,z,\varphi_E,\varphi_M,\varphi_{M_a},\varphi_{M_p},\varphi_{M_s})\\
&+\Omega_E I_E+\Omega_M J_M+\Omega_{M_a} J_{M_a}
+\Omega_{M_p} J_{M_p}+\Omega_{M_s} J_{M_s}~~.
\end{aligned}
\end{equation}
It is immediate to verify that Hamilton's equations for all the eight pairs of canonically conjugate variables of the Hamiltonian (\ref{eq:HamSphExt}) yield the same equations of motion for the three coordinates $(\rho,\Phi,z)$ and momenta $(p_\rho,p_\Phi,p_z)$ as the (time-dependent) ones of the original Hamiltonian (\ref{eq:HamSphPre}). However, the formal appearance of the new Hamiltonian (\ref{eq:HamSphExt}) as autonomous allows to control the accuracy, e.g., of numerical integrations of the flow of (\ref{eq:HamSphExt}) by checking, e.g., the error in energy preservation. Furthermore, since the geopotential part of $V$ depends only on the combination $\varphi=\Phi-\varphi_E$, it turns out convenient to define a pair of canonically conjugate action-angle variables $(\varphi,p_\varphi)$ via the canonical transformation $\Phi=\varphi+\varphi_E$, $p_\Phi=p_\varphi$, $I_E=J_E-p_\varphi$. The Hamiltonian then takes the form
\begin{equation}\label{eq:HamSphe}
\begin{aligned}
&H\equiv H(\rho,\varphi,z,\varphi_E,\varphi_M,\varphi_{M_a},\varphi_{M_p},\varphi_{M_s},p_\rho,p_\Phi,p_z,J_E,J_M,J_{M_a},J_{M_p},J_{M_s})=\\
&=\frac{{p_\rho}^2}{2}+\frac{{p_\varphi}^2}{2\rho^2}+\frac{p_z^2}{2} - \Omega_E p_\varphi
+V(\rho,\varphi,z,\varphi_E,\varphi_M,\varphi_{M_a},\varphi_{M_p},\varphi_{M_s})\\
&+\Omega_E J_E+\Omega_M J_M+\Omega_{M_a} J_{M_a}
+\Omega_{M_p} J_{M_p}+\Omega_{M_s} J_{M_s}~~.
\end{aligned}
\end{equation}
The Hamiltonian \eqref{eq:HamSphe} is the starting point of our analytical study of GEO long-term dynamics, via the appropriate construction of a normal form.

%%%%%%%%%%%%%%%%%%%%%%%%%%%%%%%%%%%%%%%%%%%%%%%%%%%%%%%%%%%%%%%%%%%%%%%%%%%%%%%%%%%%%%%%%
\section{Normal form}
\label{sec:hamexp}
%%%%%%%%%%%%%%%%%%%%%%%%%%%%%%%%%%%%%%%%%%%%%%%%%%%%%%%%%%%%%%%%%%%%%%%%%%%%%%%%%%%%%%%%%
%------------------------------------------------
\subsection{Hamiltonian expansions}
%------------------------------------------------
The model \eqref{eq:HamSphe} already contains some approximations related to the use of only some terms in the geopotential, the choice of lunisolar orbit models, the omission of the Earth's shadowing effect on the SRP, etc.
Despite these simplifications, the model \eqref{eq:HamSphe} is still quite complicated for analytical investigations. We now introduce some further expansions of the Hamiltonian \eqref{eq:HamSphe} in order to bring it to a form suitable for the construction of a normal form. In sequence, these expansions are described as follows:

%---------------------------------------------------------------------------
\subsubsection{Expansion of the lunisolar and SRP potentials: book-keeping } 
%---------------------------------------------------------------------------
\label{sec:bookk}
The potentials of the Moon \eqref{eq:potmoon}, the Sun \eqref{eq:potsun} and the SRP \eqref{eq:SRPfull} depend on the small quantities $r/r_{\odot}$ and $r/r_{\leftmoon}$.
However, the vectors $\mathbf{r_\odot}$ and $\mathbf{r_{\leftmoon}}$,  determined via Eqs. \eqref{eq:rlambdasun}, \eqref{eq:Sunelts}, \eqref{eq:rmoon}, \eqref{eq:lambdamoon}, \eqref{eq:betamoon} of Appendix A, depend themselves on a number of small parameters related to the Sun's and Moon's orbital eccentricities, inclinations, Moon's precession terms, etc.
In order to account simultaneously for {\it all} these types of small parameters in the series, we make now use of the \emph{book-keeping} technique, introduced in \citet{efthyLaPlata}.
Namely, we introduce, in all expressions, the use of a symbol $\lambda$, whose numerical value is $\lambda=1$, and whose powers $\lambda,\lambda^2,\lambda^3,\ldots$ appear as factors in front of particular groups of terms indicating that a particular group is considered as of first, second, third, etc. order of smallness.
Which power of $\lambda$ will appear in front of each group is a choice made in advance, according to the expected physical values of the various small parameters. Such a choice is called a `book-keeping' rule, representing an efficient weighting of the various perturbations coexisting in the same Hamiltonian or in the equations of motion.   

In the present series construction, we implement the book-keeping technique in three separate stages: 

i) we consider a `lunisolar' book-keeping parameter $\lambda_{ls}$, assigning, internally, orders to all the terms appearing in the expansions of the lunisolar + SRP potential terms in \eqref{eq:HamSphe}. 

ii) We consider a polynomial book-keeping parameter $\lambda_{pol}$ whose use is explained in subsection \ref{sec:epi}.

iii) Finally, we introduce the general Hamiltonian book-keeping $\lambda$ used in the normal form and all subsequent series expansions, which substitutes all previous partial book-keeping parameters.

We now give the details of the book-keeping rules for the parameter $\lambda_{ls}$. In Eqs.~\eqref{eq:potsun} and \eqref{eq:SRPfull} we set:
\begin{equation}\label{eq:bkls}
\begin{aligned}
|\mathbf{r}-\mathbf{r_\odot}|
&\longmapsto 
r_\odot\left(1-2\lambda_{ls}{\mathbf{r}\cdot\mathbf{r_\odot}\over r_\odot^2}
+\lambda_{ls}^2{r^2\over r_\odot^2}\right)^{1/2}, \
{\mathbf{r}\cdot\mathbf{r_\odot}\over r_\odot^3}
&\longmapsto 
\lambda_{ls}{\mathbf{r}\cdot\mathbf{r_\odot}\over r_\odot^3}
\end{aligned}
\end{equation}
and expand $V_\odot$ and $V_{SRP}$ up to order 2 in $\lambda_{ls}$.
A similar book-keeping assignment is made for the quantities $|\mathbf{r}-\mathbf{r_{\leftmoon}}|$ and $\mathbf{r}\cdot\mathbf{r_{\leftmoon}}/r_{\leftmoon}^3$ in $V_{\leftmoon}$, which is then expanded up to order 4 in $\lambda_{ls}$ since \citet{Musen1961b} already notes that at this altitude, at least a third order expansion is needed as $|\mathbf{r}/\mathbf{r_{\leftmoon}}| \sim 0.1$ at GEO.

These expansions contain, now, the quantities $\mathbf{r_\odot}$, $\mathbf{r_{\leftmoon}}$, which, themselves, depend on several small parameters, namely the eccentricities and inclinations of the Keplerian parts of the solar and lunar orbits, as well as the amplitudes of the lunar precession terms.
Explicit formulas for these book-keeping assignments are given in Appendix A. To reflect these choices, a new expansion up to order two with respect to the book-keeping parameter $\lambda_{ls}$ is performed.
We emphasize that the choice of book-keeping rules as above is not the only possible one.
As explained in \citet{efthyLaPlata}, different choices of book-keeping rules are possible to make, and the optimal choice can be found in practice after some experimentation.

%----------------------------------------------------------------------------
\subsubsection{Epicyclic action-angle variables. Final Hamiltonian expansion} 
%----------------------------------------------------------------------------
\label{sec:epi}
The axisymmetric part of the geopotential (Keplerian + $J_2$ terms in Eq.~\eqref{eq:VGEO2}) reads:
\begin{equation}\label{eq:VGEO2ax}
V_{GEO_0}(\rho,z)=
-\frac{\mu _\oplus}{\sqrt{\rho^2+z^2}}+\frac{\sqrt{5} \bar C_{2,0} \mu _\oplus R_\oplus^2}{2(\rho^2+z^2)^{3/2}}-\frac{3 \sqrt{5} \bar C_{2,0} \mu _\oplus R_\oplus^2 z^2}{2(\rho^2+z^2)^{5/2}}~~.
\end{equation}
The angular velocity of an equatorial circular orbit at the distance $\rho$ is given by
\begin{equation}
W(\rho)=\sqrt{\frac{1}{\rho}\frac{dV_{GEO_0}(\rho,z)}{d\rho}}\Bigg|_{z=0}~~.
\end{equation}
The radius $\rho_c$ at which $W(\rho_c)=\Omega_E$ is hereafter called the \emph{geostationary radius}. We find 
\begin{equation}
\rho_c={\mu_\oplus^{1/3}\over\Omega_E^{2/3}}
+{{\cal J}_2^2\Omega_E^{2/3}R_\oplus^2\over 2\mu_\oplus^{1/3}}
+{\cal O}({\cal J}_2^2) = 42164.69 \, \mbox{km}
\end{equation}
with ${\cal J}_2 = -\sqrt{5}\bar{C}_{2,0}$ $=$ $1.08262668355\times 10^{-3}$.
The angular momentum per unit mass of a particle in circular orbit at the geostationary radius is equal to $p_c=\Omega_E \rho_c^2$. We call effective potential the quantity
\begin{equation}
\label{eq:Veff}
V_{GEO_{eff}}=\frac{p_c^2}{2\rho^2}+V_{GEO_0}(\rho,z)~~.
\end{equation}
The effective potential describes the epicyclic oscillations of particles in nearly circular orbits under the axisymmetric potential $V_{GEO_0}(\rho,z)$, with (preserved) value of the $z$-component of the angular momentum $p_\varphi=p_c$. Setting $\rho=\rho_c+\delta\rho$,  expanding $V_{GEO_{eff}}$ up to terms of second degree in $\delta\rho$ and $z$, and using the equation of the circular orbit, we find the harmonic potential terms
\begin{equation}
V_{GEO_{eff}}=const + {1\over 2}\kappa^2\delta\rho^2 + {1\over 2}\kappa_z^2 z^2 +\ldots
\end{equation}
where
\begin{equation}\label{eq:kappaep}
\kappa = \Omega_E
-{3{\cal J}_2\Omega_E^{7/3}R_\oplus^2\over 2\mu_\oplus^{2/3}}+{\cal O}({\cal J}_2^2)~,~~~~
\kappa_z = \Omega_E
+{3{\cal J}_2\Omega_E^{7/3}R_\oplus^2\over 2\mu_\oplus^{2/3}}+{\cal O}({\cal J}_2^2)~.
\end{equation}
The quantities $\kappa$ and $\kappa_z$ are called the radial and vertical epicyclic frequencies respectively. We find $\kappa=6.300154$~rad~day\textsuperscript{-1}, $\kappa_z=6.300622$~rad~day\textsuperscript{-1}.

The epicyclic motion refers to the time variations of $\delta\rho(t)$, $z(t)$, both quantities considered small with respect to $\rho_c$. In the harmonic approximation, these can be written as
\begin{equation}\label{eq:epiosc}
\delta\rho(t)=\delta r_0\cos[\kappa(t-t_0)+\pi], \
z(t)=\delta z_0\sin[\kappa_z(t-t_0)+\omega]+...
\end{equation}
where $t_0$ is the time of pericentric passage and $\omega$ the argument of the perigee. 
By analyzing the relation between Cartesian position variables and elements (see \citet{MurrayDermott1999}, p.51) we find (for a Keplerian ellipse with instantaneous semi-major axis $a=\rho_c$, eccentricity $e$, inclination $I$, arguments of the perigee and of the node $\omega$ and $\Omega$, and true anomaly $f$):
\begin{equation}\label{eq:epiele}
\rho=a-a e\cos M+ O_2,~~~ z=a\sin I\sin(M+\omega)+ O_2,~~~\Phi=f+\omega+\Omega+O_2,
\end{equation}
where $M$ is the mean anomaly and $O_j$ means terms of $j^{th}$ order in the eccentricity and inclination. Comparing Eqs.(\ref{eq:epiosc}) and (\ref{eq:epiele}), we find the correspondence $\delta\rho\simeq a e=\rho_c e$, $\delta z_0\simeq a\sin I =\rho_c\sin I$. Also $<\dot{\Phi}>=\Omega_E\simeq<\dot{f}>+<\dot{\omega}>+<\dot{\Omega}>$. Thus 
\begin{equation}
s=<\dot{\Omega}>= \Omega_E-\kappa_z=-{3{\cal J}_2\Omega_E^{7/3}R_\oplus^2\over 2\mu_\oplus^{2/3}}+{\cal O}({\cal J}_2^2)+O_2
\end{equation}
and
\begin{equation}
g=<\dot{\Omega}>+<\dot{\omega}>= \Omega_E-\kappa={3{\cal J}_2\Omega_E^{7/3}R_\oplus^2\over 2\mu_\oplus^{2/3}}+{\cal O}({\cal J}_2^2)+O_2,
\end{equation}
which are familiar relations of the $J_2$ problem. We find $-s=g= 0.000234$~rad~day\textsuperscript{-1}, corresponding to a period of precession (of the line of nodes, or of the argument of the perigee) equal to $73.48$ years.

Another useful relation is found by differentiating the second of Eqs.(\ref{eq:epiosc}) with respect to time and squaring. Then
\begin{equation}\label{eq:jzi}
\kappa_z^2\rho_c^2\sin^2 I = (p_z^2 + \kappa_z^2 z^2) + {\cal O}({\cal J}_2)O_2 +O_3~.
\end{equation}
Eq.(\ref{eq:jzi}) will be used below in our definition of epicyclic action-angle variables.

Returning to the Hamiltonian expansion, we set $\rho=\rho_c+\delta\rho$, and use the \emph{polynomial} book-keeping $\delta\rho\longmapsto\lambda_{pol}\delta\rho$, $z\longmapsto\lambda_{pol}z$. Let $H_{exp}$ be the Hamiltonian after the expansions of subsection \eqref{sec:bookk}. We can now expand the Hamiltonian $H_{exp}$ up to some maximum degree $N_{pol}$ with respect to $\lambda_{pol}$. This gives the Hamiltonian a polynomial form in the variables $\delta\rho$ and $z$. Finally, we set $J_\varphi = p_\varphi-p_c$. Then, apart from constants, we get:
\begin{equation}
\label{hampol}
\begin{aligned}
&H_{pol}(\delta\rho,\varphi,\delta z,\varphi_E,\varphi_M,\varphi_{M_a},\varphi_{M_p},\varphi_{M_s},p_\rho,J_\varphi,p_z,J_E,J_M,J_{M_a},J_{M_p},J_{M_s};\lambda_{pol},\lambda_{ls},A/m)=\\
&\lambda_{pol}^2\left({p_\rho^2\over 2} + {p_z^2\over 2} +{1\over 2}\kappa^2\delta\rho^2+{1\over 2}\kappa_z^2z^2\right)
+\Omega_E J_E + \Omega_M J_M 
+ \Omega_{M_a} J_{M_a} + \Omega_{M_p} J_{M_p} + \Omega_{M_s} J_{M_s} \\
&+ H_{pert}(\delta \rho,\varphi,z,\varphi_E,\varphi_M,\varphi_{M_a},\varphi_{M_p},\varphi_{M_s},J_\varphi;\lambda_{pol},\lambda_{ls})
\end{aligned}
\end{equation}
with $H_{pert}$ polynomial up to the degree $N_{pol}$ in $\delta\rho$, $z$, up to second degree in $J_\varphi^2$, and trigonometric in all the angular variables.

As shown in the next subsection, the function $H_{pol}$ is a convenient starting basis for performing precise computations of perturbation theory, i.e., a high order normal form construction.
On the other hand, numerical integrations allow to specify the suitable truncation order (value of $N_{pol}$) for precise numerical propagations of the orbits, so that $H_{pol}$ can be used as a good substitute of the complete Hamiltonian of the problem.
We considered the propagation in time of the deviation between two orbits generated by an integration, for about a century, under $H_{exp}$ and under $H_{pol}$, starting with the same initial conditions close to GEO for different $A/m$ ratios.
We noticed a gain of about one order of magnitude in the precision of the integration of $H_{pol}$ with respect to $H_{exp}$ for an increment by 2 of the truncation order $N_{pol}$, the error being reduced to less than 100~m when $N_{pol}=18$.
The main source of the error comes from the expansion of the factors $[(\rho_c+\delta\rho)^2+z^2]^{-1/2}$ and $[(\rho_c+\delta\rho)^2+z^2]^{-3/2}$ in the Hamiltonian, due to the geopotential.
In fact, as explained in detail in the next subsection, the orientation of the Laplace plane at the geostationary distance implies that orbits started initially close to the equator can reach inclinations as high as $\sim 20^{\circ}$, or $z\sim 0.3\rho_c\approx 1.2\times 10^4$~km above or below the equator.
These numbers increase even more if $A/m$ is large.
Thus, even for circular orbits, we need to keep terms of high degree in the Hamiltonian expansion, in order to accurately represent the vertical motion of the test particle (in fact, these expansions become singular at the inclination $i=45^\circ$).

With $N_{pol}=18$, the Hamiltonian $H_{pol}$ is a sum of about 2800 Poisson-trigonometric monomials, representing terms depending on various small parameters, namely: the epicyclic variables, $\mu_\odot$, $\mu_{\leftmoon}$, $C_{2,2}$ and $S_{2,2}$, $P_r$, the terms corresponding to the eccentricity and inclination of the orbit of the Moon and the Sun, the terms corresponding to the precession of the argument of periapse and longitude of the ascending node of the Moon. Based on their expected numerical values, we now arrange the Hamiltonian terms in groups of different order of smallness. To this end, we assign to all terms powers of a general book-keeping parameter $\lambda$, by the following rule: for every factor of the form $\lambda_{pol}^{s_1} J_\varphi^{s_2}P_r^{s_3}\mu_\odot^{s_4}\mu_{\leftmoon}^{s_5}\lambda_{ls}^{s_6}\lambda_{22}^{s_7}$ in front of one term of the Hamiltonian (where $\lambda_{22}$ stands for either $\bar{C}_{22}$ or $\bar{S}_{22}$), we change this factor to
$$
\lambda_{pol}^{s_1} J_\varphi^{s_2} 
P_r^{s_3}\mu_\odot^{s_4}\mu_{\leftmoon}^{s_5}\lambda_{ls}^{s_6}\lambda_{22}^{s_7} 
\longmapsto 
\lambda^{\max\{s_1+2s_2+3s_3+3s_4+3s_5+s_6+4s_7-2,0\}}
J_\varphi^{s_2} P_r^{s_3}\mu_\odot^{s_4}\mu_{\leftmoon}^{s_5}\lambda_{22}^{s_7} ~~.
$$
Thus, $\lambda_{pol}$ and $\lambda_{ls}$ are no longer present, and, instead, the Hamiltonian is now `book-kept' only by the parameter $\lambda$, i.e., it takes the form
\begin{equation}\label{eq:hambk}
H=H_0 + \lambda H_1 + \lambda^2 H_2 + ...
\end{equation}
with
\begin{equation}
H_0 ={p_\rho^2\over 2} + {p_z^2\over 2} +{1\over 2}\kappa^2\delta\rho^2+{1\over 2}\kappa_z^2z^2 +\Omega_E J_E + \Omega_M J_M 
+ \Omega_{M_a} J_{M_a} + \Omega_{M_p} J_{M_p} + \Omega_{M_s} J_{M_s}~~.
\end{equation}
The form of the Hamiltonian up to second order in the book-keeping $\lambda$ is given in electronic form as a supplementary material along with its full development (which contains terms up to degree 20 in $\lambda$, corresponding to $N_{pol}=18$) in the form of tables.
This can be used as a substitute to a complete model for numerical propagation of orbits.

In the final step, we define the radial and vertical epicyclic action-angle variables $(J_\rho,\varphi_\rho)$, $(J_z,\varphi_z)$ via
\begin{equation}
\label{eq:EpiChange}
\begin{aligned}
\delta \rho &= \sqrt{\frac{2 J_\rho}{\kappa}} \sin \left(\varphi_\rho\right),~~~~
z = \sqrt{\frac{2 J_z}{\kappa_z}} \sin \left(\varphi_z\right),\\
p_{\rho } &= \sqrt{2 \kappa J_\rho} \cos \left(\varphi_\rho\right),~~~~
p_z = \sqrt{2 J_z \kappa_z} \cos \left(\varphi_z\right)~.
\end{aligned}
\end{equation}
The book-kept Hamiltonian is now fully expressed in action-angle variables, and it contains about 1800 trigonometric monomials. We denote this Hamiltonian as $H^{(0)}$, where the superscript $(0)$ means the initial Hamiltonian, i.e., before any canonical normalization. It is a function depending on eight canonical pairs of variables, which has the form:
\begin{equation}
\label{eq:ham0}
H^{(0)}(\boldsymbol{\varphi},\boldsymbol{J})=Z_0(\boldsymbol{J}) 
+ \lambda H^{(0)}_1(\boldsymbol{\varphi},\boldsymbol{J})
+ \lambda^2 H^{(0)}_2(\boldsymbol{\varphi},\boldsymbol{J})+\ldots
\end{equation}
where $\boldsymbol{\varphi}\equiv(\varphi_\rho,\varphi,\varphi_z,\varphi_E,\varphi_{M},\varphi_{M_a},\varphi_{M_p},\varphi_{M_s})$, $\boldsymbol{J}\equiv(J_\rho,J_\varphi,J_z,J_E,J_{M},J_{M_a},J_{M_p},J_{M_s})$, with the action variables $J_E,J_M,J_{M_a}$, $J_{M_p},J_{M_s}$~being `dummy', i.e., appearing only in the unperturbed linear part of the Hamiltonian:
\begin{equation}
\label{eq:z0}
Z_0 =\kappa J_\rho + \kappa_zJ_z +\Omega_E J_E + \Omega_M J_M 
+ \Omega_{M_a} J_{M_a} + \Omega_{M_p} J_{M_p} + \Omega_{M_s} J_{M_s}~~.
\end{equation}

%------------------------------------------------------------------
\subsection{Canonical normalization via Lie series}
\label{sec:norm}
%------------------------------------------------------------------
The Hamiltonian (\ref{eq:ham0}) contains already the information about the phase space structure, particular solutions, resonances, stability, etc. of the problem under study. In order to unravel this information, the core of the canonical normalization approach is to find a transformation from the set $(\boldsymbol{\varphi},\boldsymbol{J})$ to new variables in which the dynamics becomes more transparent.
To this end, we use below the method of canonical normalization based on Lie series (see \citet{Hori1966,Deprit1969,efthyLaPlata}) and detailed in Appendix B.

%---------------------------------------------------------------------------------------
\subsection{Slow and fast variables: first normalization}
%---------------------------------------------------------------------------------------
In selecting the rules for implementing the canonical normalization of our Hamiltonian, we note immediately the main difficulty connected to the analysis of geostationary orbits, i.e., the existence of a spectrum of time scales, which renders problematic a clear-cut distinction of `fast' and `slow' variables.
To be more specific, in the Hamiltonian (\ref{eq:ham0}) there co-exist variables evolving within very different timescales, indicated by the corresponding frequencies: $\kappa$, $\kappa_z$ and $\Omega_E$ are all approximately equal to $2\pi$day$^{-1}$, due to the near-Keplerian character of the potential.
Also $\Omega_{M_a}\simeq 2\pi$month$^{-1}$, while  $\Omega_{M}= 2\pi$yr$^{-1}$,  $\Omega_{M_p}\simeq 2\pi$(8.85 yr)$^{-1}$,  $\Omega_{M_s}\simeq 2\pi$(18.6 yr)$^{-1}$.
Finally $\Omega_\varphi = 0 = 2\pi/\infty$ (geostationary condition), while, for orbits librating around the stable geostationary points, we find a libration frequency about equal to $2\pi(2.23 \text{yr})^{-1}$.
This great diversification of the timescales is an important obstacle in analytical approximations of the study of geostationary orbits.
A further obstacle is posed by the existence of secular resonances.
For example: $|\kappa-\Omega_E+\Omega_M| \simeq \Omega_M$ (the so-called `evection resonance'), but also $\Omega_{M_p}-2 \Omega_{M_s}\simeq 0$.
Such resonances may introduce weak chaos effects, thus limiting the precision of normal form computations. 

In the sequel we present results obtained by a practical splitting of the normalization procedure in two stages. In the first stage, we eliminate from the Hamiltonian the terms depending on `fast' angles, the latter being arbitrarily defined as angles with frequencies greater or equal to the monthly one $\Omega_{M_a}$. After some experimentation, we found that this arbitrary threshold separating `fast' from `slow' degrees of freedom in the Hamiltonian turns to be convenient from an algorithmic point of view, while allowing to obtain reasonably good analytical approximations to the dynamics. The normal form produced after the first stage represents a `secular' Hamiltonian, in the sense that all trigonometric terms in this Hamiltonian depend on angles with periods of the order of one year or larger, i.e., a factor $10^2$ or $10^3$ larger than the most basic (i.e., the daily) period, and a factor at least 12 larger than the monthly period. In the second stage, we perform a further canonical normalization of the secular Hamiltonian, aiming to obtain more detailed information in the space of  the secular variables. In particular, we define with greater accuracy the low-dimensional torus solution corresponding to the position of the forced equilibrium in the space of secular variables. Also, we indicate below how to approximately compute `proper elements' (i.e., approximate integrals of motion) derived from the secular variables.

More specifically: at the first normalization stage we consider the set of integers quadruplets
$(k_1,k_3,k_4,k_6)\in{\mathbb{Z}}^4$, defined by 
\begin{equation}
\label{eq:resmod}
{\cal M} = \left\{(k_1,k_3,k_4,k_6): k_1+k_3+k_4=0~\mbox{and}~k_6=0\right\}~~.
\end{equation}
At every normalization step $r=1,2,\ldots$ the corresponding Hamiltonian term $H^{(r-1)}_r$ contains terms of the form 
$$
a_{\boldsymbol{s},\boldsymbol{k}} J_\rho^{s_1}J_\varphi^{s_2}J_z^{s_3}
\exp i(k_1\varphi_\rho+k_2\varphi+k_3\varphi_z+k_4\varphi_E+
k_5\varphi_M+k_6\varphi_{M_a}+k_7\varphi_{M_p}+k_8\varphi_{M_s})
$$
Then, the term $h^{(r-1)}_r$ of Eq.(\ref{eq:homo}) in Appendix B is formed by all the terms of $H^{(r-1)}_r$ for which  $(k_1,k_3,k_4,k_6)\notin{\cal M}$. Conversely, the normal form contains terms for which 
$(k_1,k_3,k_4,k_6)\in{\cal M}$.\\
The normal form obtained by taking $N_{pol}=8$ and normalizing up to order $r=6$ is given in supplementary material.
The qualitative features of the secular dynamics are determined by a particular subset of all the terms which appear in the normal form. The most important terms appear up to book-keeping order 5 (terms of order 1 are only constants, thus they can be omitted from the normal form). These terms are the following:\\
\\
\noindent{Order 0:}
$$
\kappa J_\rho + \kappa_zJ_z +\Omega_E J_E + \Omega_M J_M 
+ \Omega_{M_a} J_{M_a} + \Omega_{M_p} J_{M_p} + \Omega_{M_s} J_{M_s}~~.
$$
\noindent{Order 2:}
$$
\left({1\over\sqrt{\kappa}}+{\sqrt{\kappa}\over 2(\Omega_E-\Omega_M)}\right)
P_rC_r{A\over m}\sqrt{2J_\rho}\sin(\varphi_\rho-\varphi-\varphi_E+\varphi_M)
$$
$$
-{3(J_\varphi+J_\rho+J_z)^2\over 2\rho_c^2}
-{\sqrt{15}\kappa^2R_E^2\over 2}
\left(\overline{C}_{2,2}\cos(2\varphi)+\overline{S}_{2,2}\sin(2\varphi)\right)~~.
$$
\noindent{Order 3:}
$$
-{7\over 4\kappa}\left({\mu_\odot\over a_\odot^3}+{\mu_{\leftmoon}\over a_{\leftmoon}^3}\right)J_\rho
-{1\over 4\kappa_z}\left({\mu_\odot\over a_\odot^3}+{\mu_{\leftmoon}\over a_{\leftmoon}^3}\right)J_z
-\frac{\Omega_E}{\kappa^2}\left({\mu_\odot\over a_\odot^3}+{\mu_{\leftmoon}\over a_{\leftmoon}^3}\right)J_\varphi
$$
$$
-{3\sin 2\varepsilon\rho_c\over 8\sqrt{\kappa_z}}
\left({\mu_\odot\over a_\odot^3}+{\mu_{\leftmoon}\over a_{\leftmoon}^3}\right)
\sqrt{2J_z}
\bigg(\sin(\Omega_G)\sin(\varphi_z-\varphi-\varphi_E)+\cos(\Omega_G)\cos(\varphi_z-\varphi-\varphi_E)\bigg)~~.
$$
\noindent{Order 4:}
$$
{2(J_\rho+J_\varphi+J_z)^3\over \kappa\rho_c^4}
+\left(\frac{11}{2} J_\rho+3J_\varphi+\frac{7}{2} J_z\right)\frac{\sqrt{15}\kappa R_E^2}{\rho_c^2}
\left(\overline{C}_{2,2}\cos(2\varphi)+\overline{S}_{2,2}\sin(2\varphi)\right)
$$
$$
+{\left(\frac{15}{2}J_\rho+9J_\varphi+6 J_z\right)\over 4\rho_c^2(\Omega_E-\Omega_M)\sqrt{\kappa}}{A\over m}P_rC_r 
\sqrt{2J_\rho}\sin(\varphi_\rho-\varphi-\varphi_E+\varphi_M)
$$
$$
+{3\mu_EC_rP_r\sin(\varepsilon)\over 8\rho_c^4\kappa_z\sqrt{\kappa\kappa_z}}
\left({1\over \kappa}+{1\over 2\kappa_z-\kappa}\right)\left({A\over m}\right)
\sqrt{2J_{\rho}}\sqrt{2J_{z}}~~~~~~~~~~~~~~~~~~~~~~~~~~~~~
$$
$$
~~~~~~~~~~~~~~~~~~~~\times
\bigg(\sin(\Omega_G)\cos(\varphi_{\rho}-\varphi_{z}+\varphi_M)
+\cos(\Omega_G)\sin(\varphi_{\rho}-\varphi_{z}+\varphi_M)\bigg)~~.\\
$$
\noindent{Order 5:}
$$
\left(-\frac{21e_{\odot}^2\mu_{\odot}}{8a_{\odot}^3\kappa}-\frac{21e_{\leftmoon}^2\mu_{\leftmoon}}{8a_{\leftmoon}^3\kappa}+{11\sin^2\varepsilon\over 4\kappa}\left({\mu_\odot\over a_\odot^3}+{\mu_{\leftmoon}\over a_{\leftmoon}^3}\right)\right)J_\rho
$$
$$
\left(-\frac{3e_{\odot}^2\mu_{\odot}}{8a_{\odot}^3\kappa_z}-\frac{3e_{\leftmoon}^2\mu_{\leftmoon}}{8a_{\leftmoon}^3\kappa_z}+{7\sin^2\varepsilon\over 16\kappa_z}\left({\mu_\odot\over a_\odot^3}+{\mu_{\leftmoon}\over a_{\leftmoon}^3}\right)\right)J_z
$$
$$
\left(-\frac{3\Omega_E e_{\odot}^2\mu_{\odot}}{8a_{\odot}^3\kappa^2}-\frac{3\Omega_E e_{\leftmoon}^2\mu_{\leftmoon}}{8a_{\leftmoon}^3\kappa^2}+\frac{13\sin^2\varepsilon\Omega_E }{8 \kappa^2}\left({\mu_\odot\over a_\odot^3}+{\mu_{\leftmoon}\over a_{\leftmoon}^3}\right)\right)J_\varphi
$$
The appearance of the angles $\varphi_\rho-\varphi-\varphi_E+\varphi_M$ and $\varphi_z-\varphi-\varphi_E$ motivates the following canonical change of variables:\\
\\
\begin{minipage}{.45\linewidth}
\begin{equation}
\nonumber
\begin{aligned}
\varphi_{ec}&=\varphi_\rho-\varphi-\varphi_E+\varphi_M,\\
\varphi_R&=\varphi,\\
\varphi_{in}&=\varphi_z-\varphi-\varphi_E,\\
\varphi_e&=\varphi_E,\\
\varphi_m&=\varphi_M,\\
\varphi_{ma}&=\varphi_{Ma},\\
\varphi_{mp}&=\varphi_{Mp},\\
\varphi_{ms}&=\varphi_{Ms},
\end{aligned}
\end{equation}
\end{minipage}
\begin{minipage}{.45\linewidth}
\begin{equation}
\label{eq:sfang}
\begin{aligned}
J_{ec}&=J_\rho,\\
J_R&=J_\varphi+J_\rho+J_z,\\
J_{in}&=J_z,\\
J_e&=J_E+J_\rho+J_z,\\
J_m&=J_{\varphi_M}-J_\rho,\\
J_{ma}&=J_{\varphi_{Ma}},\\
J_{mp}&=J_{\varphi_{Mp}},\\
J_{ms}&=J_{\varphi_{Ms}}.
\end{aligned}
\end{equation}
\end{minipage}\\
\\
Substituting the transformation (\ref{eq:sfang}) within all the normal form terms we are led to a normal form that can be decomposed as:
\begin{equation}
\label{eq:Zdecomp}
Z=Z_{sec}+Z_{res}+Z_{rest}
\end{equation}
where i) $Z_{sec}$ contains terms depending only on the canonical pairs $(\varphi_{ec},J_{ec})$ and $(\varphi_{in},J_{in})$, ii) $Z_{res}$ contains terms depending either on $(\varphi_R,J_R)$ alone, or together with $(\varphi_{ec},J_{ec})$ or $(\varphi_{in},J_{in})$, iii) $Z_{rest}$ contains all the remaining terms, depending on the slow angles, combinations of $(\varphi_M,\varphi_{M_a},\varphi_{M_p},\varphi_{M_s})$.
The meaning of the decomposition in \eqref{eq:Zdecomp} can be better understood by considering only a subset of the terms of $Z_{sec}$ and $Z_{res}$, called $Z_{sec,simple}$ and $Z_{res,simple}$ respectively.
This defines a simplified Hamiltonian $Z_{simple}=Z_{sec,simple}+Z_{res,simple}$.
Restoring the numerical value of the book-keeping parameter $\lambda=1$, the simplified model deduced from the secular normal form, which contains the most important terms appearing in $Z$, reads (simplifying $\Omega_E \approx \kappa \approx \kappa_z$ where necessary):
\begin{equation}
\label{eq:zsimpledef}
Z_{simple}=Z_{sec,simple}+Z_{res,simple}
\end{equation}
where
\begin{eqnarray}\label{eq:zsimple}
Z_{sec,simple} 
~&= &
\left[
\Omega_M-{3{\cal J}_2\Omega_E^{7/3}R_E^2\over 2\mu_E^{2/3}}
-\left({3\over 4\kappa}\left(1+\frac{3}{2}\left(e_{\odot}^2-\sin^2\varepsilon\right)\right)\right)
{\mu_\odot\over a_\odot^3}-\left({3\over 4\kappa}\left(1+\frac{3}{2}\left(e_{\leftmoon}^2-\sin^2\varepsilon\right)\right)\right){\mu_{\leftmoon}\over a_{\leftmoon}^3}\right.\nonumber\\
~&- &
\left.{3P_rC_r \over 8\rho_c^2(\Omega_E-\Omega_M)\sqrt{\kappa}}{A\over m}
\sqrt{2J_{ec}}\sin(\varphi_{ec})
\right] J_{ec} \nonumber\\
~&+&\left({1\over\sqrt{\kappa}}+{\sqrt{\kappa}\over 2(\Omega_E-\Omega_M)}\right)
P_rC_r{A\over m}\sqrt{2J_{ec}}\sin(\varphi_{ec})\nonumber\\
~&+ &
\left[
{3{\cal J}_2\Omega_E^{7/3}R_E^2\over 2\mu_E^{2/3}}
+\left({3\over 4\kappa}\left(1+\frac{3}{2}e_{\odot}^2\right)-\frac{19}{16\kappa}\sin^2\varepsilon\right)
{\mu_\odot\over a_\odot^3}+\left({3\over 4\kappa}\left(1+\frac{3}{2}e_{\odot}^2\right)-\frac{19}{16\kappa}\sin^2\varepsilon\right){\mu_{\leftmoon}\over a_{\leftmoon}^3}\right.\nonumber\\
~&- &
\left.{3P_rC_r \over 4\rho_c^2(\Omega_E-\Omega_M)\sqrt{\kappa}}{A\over m}
\sqrt{2J_{ec}}\sin(\varphi_{ec})
\right] J_{in} \nonumber\\
~&- &
{3\sin (2\varepsilon)\rho_c\over 8\sqrt{\kappa_z}}
\left({\mu_\odot\over a_\odot^3}+{\mu_{\leftmoon}\over a_{\leftmoon}^3}\right)
\sqrt{2J_{in}}
\bigg(\sin(\Omega_G)\sin(\varphi_{in})+\cos(\Omega_G)\cos(\varphi_{in})\bigg)\nonumber\\
~&+ &
{3\mu_EC_rP_r\sin(\varepsilon)\over 8\rho_c^4\kappa_z\sqrt{\kappa\kappa_z}}
\left({1\over \kappa}+{1\over 2\kappa_z-\kappa}\right)\left({A\over m}\right)
\sqrt{2J_{ec}}\sqrt{2J_{in}}\nonumber\\
~&~&
~~~~~~~~~~~~~~~~~~\times\bigg(\sin(\Omega_G)\cos(\varphi_{ec}-\varphi_{in})
+\cos(\Omega_G)\sin(\varphi_{ec}-\varphi_{in})\bigg),
\end{eqnarray}
\begin{eqnarray}\label{eq:zressimple}
Z_{res,simple} 
~&= &-\frac{1}{\kappa}\left({\mu_\odot\over a_\odot^3}\left(1+\frac{3e_\odot^2}{2}-\frac{13\sin^2\varepsilon}{8}\right)
+{\mu_{\leftmoon}\over a_{\leftmoon}^3}\left(1+\frac{3e_{\leftmoon}^2}{2}-\frac{13\sin^2\varepsilon}{8}\right)\right)J_R-{3J_R^2 \over 2\rho_c^2}+{2J_R^3\over \kappa\rho_c^4}\nonumber
\\
~&+ &
{9P_rC_r J_R\over 4\rho_c^2(\Omega_E-\Omega_M)\sqrt{\kappa}}{A\over m}
\sqrt{2J_{ec}}\sin(\varphi_{ec})-{\sqrt{15}\kappa^2 R_E^2\over 2}
\left(\overline{C}_{2,2}\cos(2\varphi_R)+\overline{S}_{2,2}\sin(2\varphi_R)\right)\nonumber\\
~&+ &
\left(\frac{5}{2} J_{ec}+3J_R+\frac{1}{2}J_{in} \right){\sqrt{15}\kappa R_E^2\over \rho_c^2}
\left(\overline{C}_{2,2}\cos(2\varphi_R)+\overline{S}_{2,2}\sin(2\varphi_R)\right).
\end{eqnarray}
The approximative Hamiltonian model of Eqs.~\eqref{eq:zsimpledef}~--~\eqref{eq:zressimple} contains the basic features of secular dynamics in the resonant GEO domain. We now discuss the features of $Z_{simple}$.

%-------------------------------------------------------------------------------
\subsection{Secular dynamics: forced equilibrium, resonance and proper elements}
\label{sec:secdyn}
%-------------------------------------------------------------------------------
Consider the canonical change of variables:
\begin{equation}
\label{eq:poinc}
\begin{aligned}
x_e&=\sqrt{2J_{ec}}\sin(\varphi_{ec}),\\
y_e&=\sqrt{2J_{ec}}\cos(\varphi_{ec}),\\
x_i&=\sqrt{2J_{in}}\sin(\varphi_{in}),\\
y_i&=\sqrt{2J_{in}}\cos(\varphi_{in})~~.
\end{aligned}
\end{equation}
In the jargon of Celestial Mechanics, the variables $(x_e,y_e)$ and $(x_i,y_i)$ are called `Poincar\'{e}' variables; for reasons explained below, their respective planes will be hereafter called the eccentricity and inclination plane. The expression of $Z$ after this change of variables is detailed in supplementary material and the full form of $Z_{sec}$ is listed therein.

Substituting Eqs.(\ref{eq:poinc}) in the normal form $Z$, the part $Z_{sec}$ becomes a function of the canonical variables $(x_e,x_i,y_e,y_i)$. We call `forced equlibrium' a stable equilibrium solution of the equations of motion produced by the secular normal form depending only on the secular variables $(x_e,x_i,y_e,y_i)$, namely:
\begin{equation}
\label{eq:forceq}
\begin{aligned}
\dot{x}_e={\partial Z_{sec}\over\partial y_e}=0,~~~
\dot{x}_i={\partial Z_{sec}\over\partial y_i}=0,~~~\\
\dot{y}_e=-{\partial Z_{sec}\over\partial x_e}=0,~~~
\dot{y}_i=-{\partial Z_{sec}\over\partial x_i}=0.
\end{aligned}
\end{equation}
Using $Z_{sec,simple}$, approximate formulas for the GEO forced equilibrium can be found as follows:\\
\\
\noindent
i) {\em Eccentricity plane:} Ignoring higher order and coupling terms, the first two lines in Eq.(\ref{eq:zsimple}) yield the equilibrium position:
\begin{equation}\label{eq:xeforced}
x_{e,f}=-
\frac{\left({1\over\sqrt{\kappa}}+{\sqrt{\kappa}\over 2(\Omega_E-\Omega_M)}\right)
P_rC_r{A\over m}}
{\Omega_M-{3{\cal J}_2\Omega_E^{7/3}R_E^2\over 2\mu_E^{2/3}}
-\left({3\over 4\kappa}\left(1+\frac{3}{2}\left(e_{\odot}^2-\sin^2\varepsilon\right)\right)\right)
{\mu_\odot\over a_\odot^3}-\left({3\over 4\kappa}\left(1+\frac{3}{2}\left(e_{\leftmoon}^2-\sin^2\varepsilon\right)\right)\right){\mu_{\leftmoon}\over a_{\leftmoon}^3}},~~~
y_{e,f}=0
\end{equation}
In the same approximation, the motion around $(x_{e,f},y_{e,f})$ is a harmonic oscillation with a frequency:
\begin{equation}\label{eq:omee}
\Omega_{e,f}=\Omega_M-{3{\cal J}_2\Omega_E^{7/3}R_E^2\over 2\mu_E^{2/3}}
-\left({3\over 4\kappa}\left(1+\frac{3}{2}\left(e_{\odot}^2-\sin^2\varepsilon\right)\right)\right)
{\mu_\odot\over a_\odot^3}-\left({3\over 4\kappa}\left(1+\frac{3}{2}\left(e_{\leftmoon}^2-\sin^2\varepsilon\right)\right)\right){\mu_{\leftmoon}\over a_{\leftmoon}^3}
\end{equation}
Note that $\Omega_{e,f}$ does not depend on $A/m$, consistently with \citet{Rosengren2013,Valk2008a} and is equal to approximately one year.\\
The motion with initial condition $x_e(0),y_e(0)$ is given by:
\begin{equation}\label{eq:xetime}
x_e(t)=x_{e,f}+A_e\cos(\Omega_{e,f}t+\delta\phi_e),~~~ 
y_e(t)=-A_e\sin(\Omega_{e,f}t+\delta\phi_e),~~~ 
\end{equation}
with $A_e=((x_e(0)-x_{e,f})^2+y_e(0)^2)^{1/2}$, $\delta\phi_e=\arctan(x_e(0)-x_{e,f},y_e(0))$. According to Eq.(\ref{eq:xetime}) the quantity
\begin{equation}\label{eq:xeprop}
A_e =  [(x_e(t)-x_{e,f})^2+y_e(t)^2]^{1/2}
\end{equation}
is a constant of motion, or {\it proper element} expressing the distance of the orbit from the forced equilibrium in the plane $(x_e,y_e)$. The proper element can also be expressed in terms of an action variable introduced via the canonical transformation $(x_e,y_e)\rightarrow (\theta_e,I_e)$ given by:
\begin{equation}\label{eq:aaeprop}
x_e = x_{e,f}+ \sqrt{2I_e}\sin\theta_e,~~~ y_e = y_{e,f}+ \sqrt{2I_e}\cos\theta_e~.
\end{equation}
Furthermore, according to Eq.(\ref{eq:epiele}), we have
\begin{equation}\label{eq:xeecc}
x_e^2+y_e^2 = (\mu_E\rho_c)^{1/2}e^2 + O(e^3)~,
\end{equation}
whence the characterization of the plane $(x_e,y_e)$ as the `eccentricity plane'. In particular, the forced solution (\ref{eq:xeforced}) corresponds to a forced eccentricity, which is non-zero for $A/m\neq 0$. Substituting numerical values to the parameters of Eq.(\ref{eq:xeforced}), we find:
\begin{equation}
e_{forced}\approx 0.0114 {A\over m}~,
\end{equation}
consistent with \citet{Valk2008a,ChaoBaker}.
On the other hand, the angle $\varphi_{ec}$ for the forced equilibrium has a constant value $\varphi_{ec,f}=3\pi/2$ (since $x_e<0,y_e=0$). Using Eqs.(\ref{eq:sfang}) and (\ref{eq:epiele}) we find:
\begin{equation}
\varphi_{ec}=-\pi/2+\varphi_M-(\omega+\Omega)+O_2.
\end{equation}
Thus, the forced value $\varphi_{ec,f}=3\pi/2$ corresponds to a argument of the perigee precessing with the yearly frequency according to
\begin{equation}
(\omega+\Omega)_f\equiv\varpi_f = \varphi_M+O_2 \; \text{mod} \; 2\pi=\Omega_M t+O_2 \; \text{mod} \; 2\pi.
\end{equation}
This precession is clearly visible in plots in the literature depicting the secular dynamics in the eccentricity plane using the usual Delaunay-like variables, or the eccentricity vector $e\cos(\omega+\Omega),e\sin(\omega+\Omega)$ (cf. figure 7 of \citet{Valk2008a} , or Fig.~1 of \citet{Rosengren2013}, using the `Milankovich elements'). This shows also the advantage of using, instead, the epicyclic variables defined as above, in which the forced equilibrium becomes a true equilibrium point in the eccentricity plane. Finally, the quantity 
\begin{equation}\label{eq:propecc}
e_p = A_e (\mu_E\rho_c)^{-1/4}
\end{equation}
defines the quasi-integral of the `proper eccentricity'. 

\begin{figure}
\centering
\begin{minipage}[t]{.48\textwidth}
\includegraphics[width=1\textwidth]{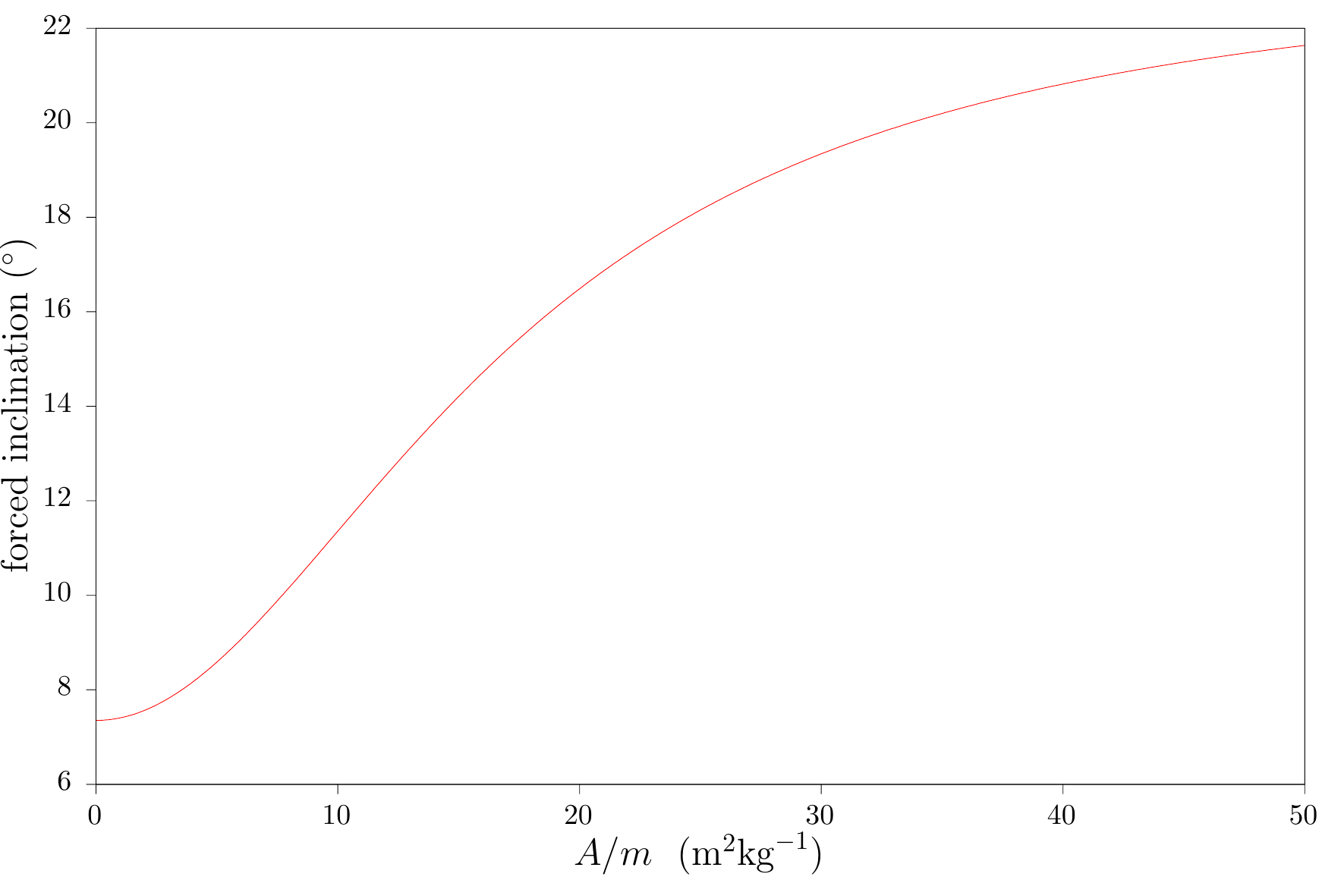}
\end{minipage}
\hfill
\begin{minipage}[t]{.48\textwidth}
\includegraphics[width=1\textwidth]{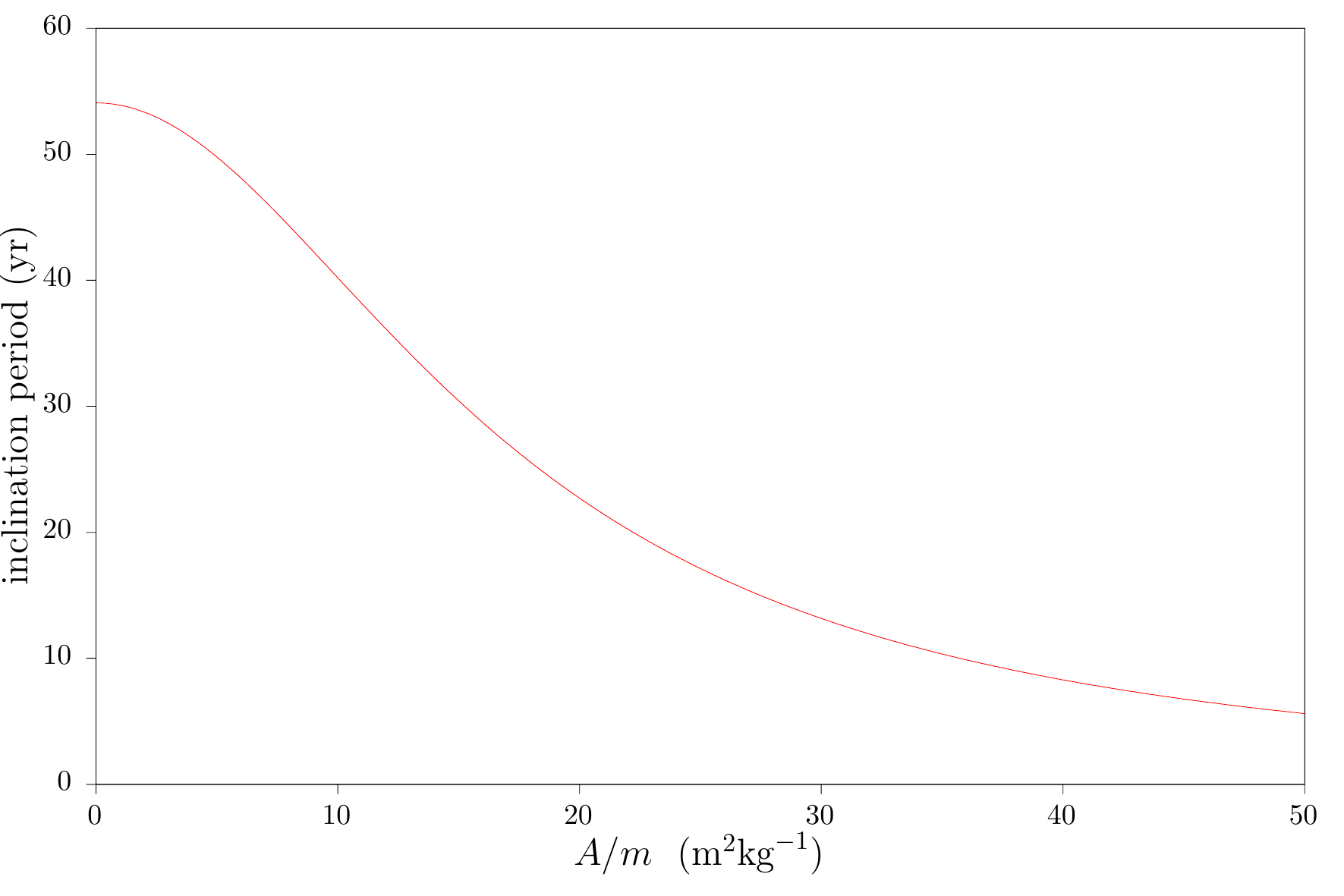}
\end{minipage}
\caption{\label{fig:incfunam}Forced inclination (left) and period of the oscillation in inclination (right) for geostationary satellites and space debris as a function of $A/m$ from Eqs.(\ref{eq:incfunam}) and \eqref{eq:omei}.}
 
\end{figure}

For space debris, a solution of particular importance is the one starting with the initial conditions $x_e(0)=y_e(0)=0$, i.e., zero initial eccentricity.
According to the mechanism proposed by \citet{Valk2008a}, space debris with a high $A/m$ may acquire a large oscillation of the eccentricity from zero to a value equal to $2e_{forced}$:
\begin{equation}\label{eq:e0time}
e(t)=2e_{forced}[1+2\cos(\Omega_{e,f}t)]^{1/2}.
\end{equation}
The resulting periodic oscillation of the eccentricity can produce a large degree of chaos near the unstable separatrix domain of the GEO resonance (see \citet{Valk2009d} and point (iii) below).\\
\\
\noindent
ii) {\em Inclination plane:} Consider the canonical change of variables:
\begin{equation}
x_i = \sin(\Omega_G)X_i-\cos(\Omega_G)Y_i,~~~
y_i = \cos(\Omega_G)X_i+\sin(\Omega_G)Y_i.
\end{equation}
Reasoning in the same way as before for the eccentricity variables, and setting $\sqrt{2J_{ec}}\sin\varphi_{ec}\equiv x_e$ equal to its forced value $x_{e,f}$, the last four lines of Eq.~\eqref{eq:zsimple} yield the following equilibrium point in the variables $(X_i,Y_i)$:
\small
\begin{equation}
\begin{aligned}
X_{i,f}&=\frac{{3\sin (2\varepsilon)\rho_c\over 8\sqrt{\kappa_z}}
\left({\mu_\odot\over a_\odot^3}+{\mu_{\leftmoon}\over a_{\leftmoon}^3}\right)
-{3\mu_EC_rP_r\sin(\varepsilon)\over 8\rho_c^4\kappa_z\sqrt{\kappa\kappa_z}}
\left({1\over \kappa}+{1\over 2\kappa_z-\kappa}\right)\left({A\over m}\right)x_{e,f}}
{{3{\cal J}_2\Omega_E^{7/3}R_E^2\over 2\mu_E^{2/3}}
+\left({3\over 4\kappa}\left(1+\frac{3}{2}e_{\odot}^2\right)-\frac{19}{16\kappa}\sin^2\varepsilon\right)
{\mu_\odot\over a_\odot^3}+\left({3\over 4\kappa}\left(1+\frac{3}{2}e_{\odot}^2\right)-\frac{19}{16\kappa}\sin^2\varepsilon\right){\mu_{\leftmoon}\over a_{\leftmoon}^3}
-{3P_rC_r \over 4\rho_c^2(\Omega_E-\Omega_M)\sqrt{\kappa}}\left({A\over m}\right)
x_{e,f}},\\
Y_{i,f}&=0.
\end{aligned}
\end{equation}
\normalsize
In view of Eq.(\ref{eq:jzi}), a non-zero value of $X_{i,f}$ implies a forced inclination
\begin{equation}\label{eq:iforced}
\sin I_{forced} = {X_{i,f}\over \sqrt{\kappa_z}\rho_c}.
\end{equation}
An orbit of inclination $I_{forced}$ lies in a plane otherwise referred to in the literature as the invariant `Laplace plane'. Eq.(\ref{eq:iforced}) provides a detailed formula for the orientation of the Laplace plane indicating its dependence on both the lunisolar gravitational perturbations and the SRP. Substituting the numerical parameter values, we find
\begin{equation}
\sin I_{forced}=\frac{0.0000407+4.355\times 10^{-7} \left(\frac{A}{m}\right)^2}{0.000318+1.0981\times 10^{-6} \left(\frac{A}{m}\right)^2},
\label{eq:incfunam}
\end{equation}
which clearly shows the dependence on  $A/m$ of the forced inclination at the geostationary radius, illustrated by the left side of Figure \ref{fig:incfunam}. This result is in agreement with the implicit expression arising from equations~6, 9, 11 and 12 in \citet{Rosengren2014b}, the difference between the two never reaching more than 0.9\%.

Regarding the orientation of the Laplace plane, in view of Eqs.(\ref{eq:sfang}) and (\ref{eq:epiele}) we find
\begin{equation}
\varphi_{in}=-\Omega + O_2.
\end{equation}
The forced solution yields $\varphi_{in}=\Omega_G$, thus $\Omega \simeq -\Omega_G$. Since $-\Omega_G$ is the angle formed between the Greenwich meridian and the equinox at 12:00 of 1 Jan JD2000, we recover the well known result that the line of nodes of the Laplace plane at GEO coincides with the Earth's Equinox.

In the harmonic approximation, the motion around $(X_{i,f},Y_{i,f})$ is a harmonic oscillation with a frequency:
\footnotesize
\begin{equation}\label{eq:omei}
\Omega_{i,f} = {3{\cal J}_2\Omega_E^{7/3}R_E^2\over 2\mu_E^{2/3}}
+\left({3\over 4\kappa}\left(1+\frac{3}{2}e_{\odot}^2\right)-\frac{19}{16\kappa}\sin^2\varepsilon\right)
{\mu_\odot\over a_\odot^3}+\left({3\over 4\kappa}\left(1+\frac{3}{2}e_{\odot}^2\right)-\frac{19}{16\kappa}\sin^2\varepsilon\right){\mu_{\leftmoon}\over a_{\leftmoon}^3}
-{3P_rC_r \over 4\rho_c^2(\Omega_E-\Omega_M)\sqrt{\kappa}}{A\over m}x_{e,f}
\end{equation}
\normalsize
This frequency is an increasing function of $A/m$ since $x_{e,f}<0$, which is consistent with the fact that the associated period decreases when $A/m$ increases, as observed in \citet{Rosengren2014b}.
The period associated to this frequency is represented on the right side of Figure \ref{fig:incfunam}.
We note that it matches quite well the analytical approximation obtained by \citep{Casanova2014}.\\
The motion with initial condition $X_i(0),Y_i(0)$ is given by:
\begin{equation}\label{eq:xitime}
X_i(t)=X_{i,f}+A_i\cos(\Omega_{i,f}t+\delta\phi_i),~~~ 
Y_i(t)=-A_i\sin(\Omega_{i,f}t+\delta\phi_i),~~~ 
\end{equation}
with $A_i=[(X_i(0)-X_{i,f})^2+Y_i(0)^2]^{1/2}$, $\delta\phi_i=\arctan(X_i(0)-X_{i,f},Y_i(0))$. The quantity
\begin{equation}\label{eq:xiprop}
A_i =  [(X_i(t)-X_{i,f})^2+Y_i(t)^2]^{1/2}
\end{equation}
is a constant of motion (or proper element) expressing the distance of the orbit from the forced equilibrium in the inclination plane $(x_i,y_i)$. This proper element can also be expressed in terms of an action variable introduced via the canonical transformation $(X_i,Y_i)\rightarrow (\theta_i,I_i)$ given by:
\begin{equation}\label{eq:aaiprop}
X_i = X_{i,f}+ \sqrt{2I_i}\sin\theta_i,~~~ Y_i = Y_{i,f}+\sqrt{2I_i}\cos\theta_i
\end{equation}
Finally, according to Eq.(\ref{eq:epiele}), we have
\begin{equation}\label{eq:xiinc}
X_i^2+Y_i^2 = x_i^2+y_i^2 = (\mu_E\rho_c)^{1/2}\sin^2 I + O_3~,
\end{equation}
whence the characterization of the plane $(x_i,y_i)$ as the `inclination plane'. In particular, the quantity 
\begin{equation}\label{eq:propinc}
I_p = A_i (\mu_E\rho_c)^{-1/4}
\end{equation}
defines the quasi-integral of the proper inclination. 

Concerning space debris, an important remark regards the fact that the frequency $\Omega_{i,f}$ (Eq.\ref{eq:omei}) depends on $(A/m)$ in such a way that for HAMR objects it can produce a low order commensurability, in particular with the lunar frequencies $\Omega_{Ms}$ (precession of the longitude of the ascending node) and $\Omega_{Mp}$ (precession of the argument of periapse). For $A/m=0$~m\textsuperscript{2}kg\textsuperscript{-1}, one has $\Omega_{Ms}\approx 3\Omega_{i,f}$, while $\Omega_{Mp}\approx 7\Omega_{i,f}$. On the other hand, due to its dependence on $x_{e,f}$, the variations of $\Omega_{i,f}$ depend quadratically on $A/m$. Thus, for $A/m>1$~m\textsuperscript{2}kg\textsuperscript{-1} one obtains significantly higher values of $\Omega_{i,f}$, resulting in lower order commensurabilities with the lunar secular frequencies. This implies that these resonances should produce weakly chaotic effects in the inclination plane, which can be unraveled via numerical simulations (see, for example \citet{Rosengren2015a} for an analogous phenomenon in the MEO region). \\
\\
\noindent
iii) {\em Resonance plane:} 
$Z_{res,simple}$ (Eq.\eqref{eq:zressimple}) refers to the dynamics in the plane $(\phi_R,J_R)$. From their definitions (Eq.(\ref{eq:sfang})), we have a) $\phi_R = \sigma+O_2$, where $\sigma = \lambda-\Omega_E t$ is the `critical argument' of the geostationary resonance, and b) $J_R = \Delta L+ O_2$, where $\Delta L = L-L_c = \sqrt{\mu_E a}-\sqrt{\mu_E \rho_c}$ is the Delaunay action associated with the orbit's instantaneous semi-major axis, measured with respect to the reference value $L_c$ at which $a=\rho_c$. 

Due to the above definitions, the separatrix structure of the geostationary resonance as depicted in the variables $(\phi_R,J_R)$ is topologically equivalent to the one found by the more familiar variables $(\sigma,\Delta L)$ (as e.g., in Figure 3 of \citet{Celletti2014}, or Figure 2 of \citet{Valk2009d}). Ignoring the coupling terms with $J_{in}$ and $J_R$ in $Z_{res,simple}$ (Eq.\eqref{eq:zressimple}), the resonant Hamiltonian is
\begin{equation}
\label{eq:zres}
\begin{aligned}
&-\frac{1}{\kappa}\left({\mu_\odot\over a_\odot^3}\left(1+\frac{3e_\odot^2}{2}-\frac{13\sin^2\varepsilon}{8}\right)
+{\mu_{\leftmoon}\over a_{\leftmoon}^3}\left(1+\frac{3e_{\leftmoon}^2}{2}-\frac{13\sin^2\varepsilon}{8}\right)\right)J_R-{3J_R^2 \over 2\rho_c^2}+{2J_R^3\over \kappa\rho_c^4}\\ 
&-{\sqrt{15}\kappa^2 R_E^2\over 2}
\left(\overline{C}_{2,2}\cos(2\varphi_R)+\overline{S}_{2,2}\sin(2\varphi_R)\right)+
{5\sqrt{15}\kappa R_E^2\over 2\rho_c^2}J_{ec}
\left(\overline{C}_{2,2}\cos(2\varphi_R)+\overline{S}_{2,2}\sin(2\varphi_R)\right).
\end{aligned}
\end{equation}
The first line in the above Hamiltonian yields a typical pendulum phase space structure. The stable points are at $J_R=0$ and $\varphi_{R1} = 1.31023$~rad, or $\varphi_{R2} = 4.45183$~rad (corresponding to a geographic longitude $75.0712^\circ \, \text{E}$ or $104.929^\circ \, \text{W}$ with respect to the Greenwich meridian). The separatrix half-width is given by:
\begin{equation}
\Delta J_R\approx 
\left({2\sqrt{15}\kappa^2 \rho_c^2R_E^2\sqrt{\overline{C}_{2,2}^2+\overline{S}_{2,2}^2}
\over 3}\right)^{1/2}.
\end{equation}
An orbit in the separatrix domain undergoes variations of its instantaneous semi-major axis of half-width
\begin{equation}
\Delta a \approx 2\sqrt{\frac{\rho_c}{\mu_E}}\Delta J_R \simeq 34\mbox{km},
\end{equation}
a value consistent with \citet{Celletti2014,Valk2009c}.
This last number practically sets the overall width of the geostationary resonance domain. On the other hand, for HAMR objects, this domain is substantially smaller, due to chaotic effects induced via the coupling of the resonant degree of freedom with the eccentricity one. The time evolution of $J_{ec}(t)$ for a HAMR object can be approximated via the temporal solution of Eq.(\ref{eq:xetime}), i.e.,
\begin{equation}
J_{ec}(t) \approx {1\over 2}\left(x_{e,f}^2 + A_e^2 +2x_{e,f}A_e\cos(\Omega_{e,f}t +\delta\phi_e)\right)
\end{equation} 
Since $x_{e,f}={\cal O}(A/m)$, we see that the SRP produces a modulated pendulum effect in the Hamiltonian \eqref{eq:zres}, of amplitude proportional to $A/m$. This effect is responsible for the production of chaos in the GEO resonance \citep{Valk2009d,Celletti2014}. In fact, chaos becomes dominant when $A/m \sim 1$~m\textsuperscript{2}kg\textsuperscript{-1}, and nearly destroys all stable motions if $A/m>50$~m\textsuperscript{2}kg\textsuperscript{-1}.

\section{Numerical results}
\label{sec:numres}
\subsection{Refined equilibrium}
The normal form $Z$ obtained after the first normalization and the recognition of its important terms via $Z_{simple}$ allowed to obtain approximate formulas for the forced equilibrium and the motion (secular or resonant) around it.
A more precise determination of the forced equilibrium can be done as follows:\\
i) solve numerically Eqs. \eqref{eq:forceq} for the variables $(x_e,x_i,y_e,y_i)$,\\
ii) the found solution $(x_{e,f},x_{i,f},y_{e,f},y_{i,f})$ is an exact equilibrium of the Hamiltonian $Z_{sec}$, but ceases to be so in the full model $Z=Z_{sec}+Z_{res}+Z_{rest}$, so we now perform a second normalization aiming to further refine the forced equilibrium solution.

\subsection{Second normalization}
\label{sec:2ndnorm}
\subsubsection{Diagonalization}
We first expand the full normal form $Z$ around $(x_{ef},y_{ef},x_{if},y_{if})$ by introducing new variables:
\begin{equation}
\label{eq:dxe}
\begin{aligned}
\delta x_e&=x_e-x_{e,f},\\
\delta y_e&=y_e-y_{e,f},\\
\delta x_i&=x_i-x_{i,f},\\
\delta y_i&=y_i-y_{i,f}.
\end{aligned}
\end{equation}
The new Hamiltonian after the transformation \eqref{eq:dxe} is expressed with coordinates representing the distance to the forced equilibrium.
We then isolate $H_{quad}$, the part of $Z_{sec}$ quadratic in $(\delta x_e,\delta y_e,\delta x_i,\delta y_i)$, and diagonalize it with respect to these variables. The diagonalization process is necessary to decouple the $(\delta x_e,\delta y_e)$ and $(\delta x_i,\delta y_i)$ variables and is done as follows:\\
We define first
\begin{equation}
\begin{aligned}
\dot{\delta x_e}={\partial H_{quad}\over\partial \delta y_e},~~~
\dot{\delta x_i}={\partial H_{quad}\over\partial \delta y_i},~~~\\
\dot{\delta y_e}=-{\partial H_{quad}\over\partial \delta x_e},~~~
\dot{\delta y_i}=-{\partial H_{quad}\over\partial \delta x_i},~~~
\end{aligned}
\end{equation}
and
\begin{equation}
\label{eq:varmat}
\begin{aligned}
\mathbf{A}= \left(
\begin{array}{cccc}
\frac{\partial \dot{\delta x_e}}{\partial \delta x_e} & \frac{\partial \dot{\delta x_e}}{\partial \delta x_i} & \frac{\partial \dot{\delta x_e}}{\partial \delta y_e} & \frac{\partial \dot{\delta x_e}}{\partial \delta y_i} \\
\frac{\partial \dot{\delta x_i}}{\partial \delta x_e} & \frac{\partial \dot{\delta x_i}}{\partial \delta x_i} & \frac{\partial \dot{\delta x_i}}{\partial \delta y_e} & \frac{\partial \dot{\delta x_i}}{\partial \delta y_i} \\
\frac{\partial \dot{\delta y_e}}{\partial \delta x_e} & \frac{\partial \dot{\delta y_e}}{\partial \delta x_i} & \frac{\partial \dot{\delta y_e}}{\partial \delta y_e} & \frac{\partial \dot{\delta y_e}}{\partial \delta y_i} \\
\frac{\partial \dot{\delta y_i}}{\partial \delta x_e} & \frac{\partial \dot{\delta y_i}}{\partial \delta x_i} & \frac{\partial \dot{\delta y_i}}{\partial \delta y_e} & \frac{\partial \dot{\delta y_i}}{\partial \delta y_i}
\end{array} \right)\ ,
\end{aligned}
\end{equation}
and calculate the eigenvalues $(\mu_1,\mu_2,\mu_3,\mu_4)$, and eigenvectors $(\mathbf{X_1},\mathbf{X_2},\mathbf{X_3},\mathbf{X_4})$ of the matrix $A$.
Let
\begin{equation}
\label{eq:amat}
\begin{aligned}
\mathbf{B}(c_1,c_2)= \left(
\begin{array}{l|l|l|l}
c_1 \mathbf{X_1} & c_2 \mathbf{X_3} & c_1 \mathbf{X_2} & c_2 \mathbf{X_4}
\end{array} \right)\ ,
\end{aligned}
\end{equation}
be a $4 \times 4$ matrix with unspecified parameters $c_1,c_2$ obtained by juxtaposing the columns $c_1 X_1$,$c_2 X_3$,$c_1 X_2$,$c_2 X_4$. We specify $c_1,c_2$ so that $\mathbf{B}$ is symplectic, i.e, that they satisfy
\begin{equation}
\label{eq:sympcon}
\begin{aligned}
\mathbf{J}=\mathbf{B}^T \mathbf{J} \mathbf{B}
\end{aligned}
\end{equation}
with
\begin{equation}
\label{eq:sympJ}
\begin{aligned}
\mathbf{J}= \left(
\begin{array}{cccc}
0 & 0 & 1 & 0\\
0 & 0 & 0 & 1\\
-1 & 0 & 0 & 0\\
0 & -1 & 0 & 0
\end{array} \right)\ .
\end{aligned}
\end{equation}
We define now $\mathbf{B_{sym}}$ as $\mathbf{B_{sym}}=\mathbf{B}(c_{1_{sym}},c_{2_{sym}})$, by which we can now introduce the following linear canonical transformation applied to the whole normal form $Z$:
\begin{equation}
\label{eq:symptrans}
\begin{aligned}
\left(
\begin{array}{c}
\delta x_e\\
\delta x_i\\
\delta y_e\\
\delta y_i
\end{array}\right)
= \mathbf{B_{sym}} \left(
\begin{array}{c}
q_e\\
q_i\\
p_e\\
p_i
\end{array}\right)\ .
\end{aligned}
\end{equation}
Finally we express $Z$ in action-angle variables via the following canonical transformation:
\begin{equation}
\label{eq:actang}
\begin{aligned}
q_e& \to \sqrt{I_e}e^{i\theta_e},\\
p_e& \to -i \sqrt{I_e}e^{-i\theta_e},\\
q_i& \to \sqrt{I_i}e^{i\theta_i},\\
p_i& \to -i \sqrt{I_i}e^{-i\theta_i}\ .
\end{aligned}
\end{equation}
Let us note that since $H_{quad}$ before the transformation was almost diagonal, the transformation \eqref{eq:symptrans} is not far from the Birkhoff notation of an identity transformation. This fact justifies keeping the subscripts $e$ and $i$ pertaining to the eccentricity and the inclination respectively. $I_e$ and $I_i$ represent in fact more precise proper elements with respect to the previously defined proper elements \eqref{eq:propecc} and \eqref{eq:propinc}.
After the change \eqref{eq:actang} we have:
\begin{equation}
H_{quad}=\Omega_{e,f} I_e + \Omega_{i,f} I_i + \Omega_E J_e + \Omega_M J_m +\Omega_{M_a} J_{ma} +\Omega_{M_p} J_{mp}+\Omega_{M_s} J_{ms}+\ldots
\end{equation}
with new frequencies $\Omega_{e,f}$ and $\Omega_{i,f}$ given by the eigenvalues of $\mathbf{A}$.
\subsubsection{Choice of module and small divisors}
The full normal form now reads:
\begin{equation}
Z=Z_{sec}(I_e,I_i,\theta_e,\theta_i)+Z_{res}(I_e,I_i,J_R,\theta_e,\theta_i,\varphi_R)+Z_{rest}(I_e,I_i,J_R,\theta_e,\theta_i,\varphi_R,\varphi_M,\varphi_{Mp},\varphi_{Ms}).
\end{equation}
By construction, all terms in $Z_{sec}$ are of order equal to or higher than one in the actions $I_e$, $I_i$. In contrast, some terms of order $O(I_e^{(1/2)})$ or $O(I_i^{(1/2)})$ exist in $Z_{res}$ and $Z_{rest}$. We now eliminate such terms by a second normalization.
To this end, we first assign new book-keeping rules to the full Hamiltonian $Z$, chosen as follows:\\
i) set the old book-keeping constant to its numerical value $\lambda=1$,\\
ii) to each term in $Z$ of the form
$$
a_{\boldsymbol{s},\boldsymbol{k}} I_e^{s_1/2}J_R^{s_2}I_i^{s_3/2}
\exp i(k_1\theta_e+k_2\varphi_R+k_3\theta_i+k_4\varphi_m+
k_5\varphi_{m_p}+k_6\varphi_{m_S}),
$$
assign a book-keeping factor
$$
\lambda^{\prime s_1+s_2+s_3-2+min(1,|k_3|+|k_4|+|k_5|+|k_6|)}.
$$
The whole normal form $Z$ is now book-kept in powers of the parameter $\lambda^{\prime}$, and it can be normalized with the same algorithm as in subsection \ref{sec:norm}.
We call $\xi_1$, $\xi_2$ the Lie generating functions performing this normalization to second order in book-keeping.
After two steps, the normalizing transformation reads:
\begin{eqnarray}
\label{eq:lietra2}
\boldsymbol{\varphi}&\equiv\boldsymbol{\varphi}^{(0)}
=&\exp(L_{\xi_{2}})\exp(L_{\xi_{1}}) \boldsymbol{\varphi}^{(2)} \\ \boldsymbol{J}&\equiv\boldsymbol{J}^{(0)}
=&\exp(L_{\xi_{2}})\exp(L_{\xi_{1}}) \boldsymbol{J}^{(2)}. \nonumber 
\end{eqnarray}
The computation of the functions $\xi_1$, $\xi_2$ is done similarly as in Eq.~\eqref{eq:solhomo} (with $\xi_r$ in place of $\chi_r$).
The functions $\xi_r$, however, are chosen so as to eliminate only terms $O(I_e^{1/2})$ or  $O(I_i^{1/2})$.
Formally, the transformation \eqref{eq:lietra2} results in that the final normal form contains terms of the form:
$$
a_{\boldsymbol{s},\boldsymbol{k}} I_e^{s_1/2}J_R^{s_2}I_i^{s_3/2}
\exp i(k_1\theta_e+k_2\varphi_R+k_3\theta_i+k_4\varphi_m+
k_5\varphi_{m_p}+k_6\varphi_{m_S})
$$
with exponents $s_1,s_2,s_3,k_1\ldots,k_6$ belonging to the set 
\begin{equation}
\label{eq:mod2}
\begin{aligned}
{\cal M}_2 &= \left\{(s_1,s_2,s_3,k_1,k_2,k_3,k_4,k_5,k_6):s_1,s_2,s_3>0, s_1+s_3\neq 1\right.\\
&\left.~~~~~~~~~~~~~~~~~~\mbox{or}~\left|k_1\Omega_{e,f}+k_3\Omega_{i,f}+k_4\Omega_M+k_5\Omega_{M_p}+k_6\Omega_{M_s}\right|\leq\Omega_{i,f}\right\}~~.
\end{aligned}
\end{equation}
The condition $s_1+s_3\neq1$ ensures that the terms in $O(I_e^{1/2})$ or  $O(I_i^{1/2})$ are eliminated and so that the new normal form contains no terms linear in $(\delta x_e,\delta y_e,\delta x_i,\delta y_i)$.
We also highlight the existence of a threshold on the magnitude of the divisors of value $\Omega_{i,f}$ in Eq.~\eqref{eq:mod2}.
The need and choice of this threshold was made empirically, after noticing that a normalization performed with no or different thresholds presented a wrong amount of secular drift in the forced equilibrium solution found by this method.
Table \ref{tab:smalldiv} shows the smallest divisors up to the threshold present in the Hamiltonian before the second normalization for $A/m=1$~m\textsuperscript{2}kg\textsuperscript{-1} (left), and $A/m=10$~m\textsuperscript{2}kg\textsuperscript{-1} (right).
The line in these tables shows where the threshold is, and the terms having a divisor in the upper part of the Tables are kept in the normal form while the terms below (and all the other ones present, but not shown here) satisfying $s_1+s_3\neq1$ are normalized and therefore not present in the normal form.
For instance, the terms in the Hamiltonian satisfying $s_1+s_3\neq1$ that have divisors present in the Tables are terms with the following divisors: $-\Omega_{e,f}+\Omega_M$, $\Omega_{i,f}$ and $-\Omega_{i,f}+\Omega_{M_s}$.
Such terms with divisors $-\Omega_{e,f}+\Omega_M$ and $\Omega_{i,f}$ are then kept in the normal form, while those with $-\Omega_{i,f}+\Omega_{M_s}$ are normalized.

\begin{table}
\footnotesize
\caption{\label{tab:smalldiv}Values of the smallest divisors and their associated period present in the Hamiltonian before the second normalization for $A/m=1$~m\textsuperscript{2}kg\textsuperscript{-1} (left) and $A/m=10$~m\textsuperscript{2}kg\textsuperscript{-1} (right.)}
\begin{tabular}[t]{ccccccccc}
\hline\hline
\rule{0pt}{3ex} divisor & value & T (yr)\\
\hline
$\Omega_{e,f}+\Omega_{i,f}-\Omega_M$ & $0.000002231$ & $7708.53$\\
$3 \Omega_{i,f}-\Omega_{M_s}$ &  $0.000063920$ & $269.12$\\
$-2 \Omega_{i,f}+\Omega_{M_s}$ & $0.000265459$ & $64.80$\\
$\Omega_{e,f}-\Omega_{i,f}-\Omega_M +\Omega_{M_s}$ & $0.000267690$ & $64.26$\\
$-2 \Omega_{e,f}-\Omega_{i,f}+2 \Omega_M$ & $0.000324916$ & $52.94$\\
$-\Omega_{e,f}+\Omega_M$ & $0.000327147$ & $52.58$\\
$\Omega_{i,f}$ & $0.000329379$ & $52.23$\\
\cline{1-3}
$\Omega_{e,f}-\Omega_{i,f}-\Omega_M+\Omega_{M_p}-\Omega_{M_s}$ & $0.000594838$ & $47.31$\\
$-\Omega_{i,f}+\Omega_{M_s}$ & $0.000594838$ & $28.92$\\
\hline\hline
\end{tabular}
\begin{tabular}[t]{ccccccccccc}
\hline\hline
\rule{0pt}{3ex} divisor & value & T (yr)\\
\hline
$2 \Omega_{e,f}+\Omega_{i,f}-2 \Omega_M$ &  $0.000009728$ & $1768.21$\\
$-2 \Omega_{i,f}+\Omega_{M_s}$ & $0.000065687$ & $261.88$\\
$-\Omega_{e,f}+2 \Omega_{i,f}+\Omega_M-\Omega_{M_s}$ & $0.000144081$ & $119.39$\\
$-\Omega_{e,f}+\Omega_M$ & $0.000209768$ & $82.01$\\
$\Omega_{e,f}+\Omega_{i,f}-\Omega_M$ & $0.000219497$ & $78.37$\\
$\Omega_{e,f}-\Omega_{i,f}-\Omega_M +\Omega_{M_s}$ & $0.000285184$ & $60.32$\\
$\Omega_{e,f}-\Omega_{i,f}-\Omega_M +\Omega_{M_p} -\Omega_{M_s}$ & $0.000381105$ & $45.14$\\
$3 \Omega_{i,f}-\Omega_{M_s}$ & $0.000363578$ & $47.31$\\
$-2 \Omega_{e,f}+2\Omega_M$ & $0.000419536$ & $41.00$\\
$\Omega_{i,f}$ & $0.000429265$ & $40.07$\\
\cline{1-3}
$-\Omega_{i,f}+\Omega_{M_s}$ & $0.000494952$ & $34.76$\\
\hline\hline
\end{tabular}
\end{table}

\subsection{Nature of the forced equilibrium and analytical expression for the original variables}
\label{sec:nat}
The sequence of transformations defined by the generating functions $\chi_r$, $r=1,\ldots 6$, and $\xi_r$, $r=1,2$, relate the original variables $(\rho,\varphi,z,p_\rho,p_\varphi,p_z)$ to the very final ones $(I_e^{(2)},I_i^{(2)},J_R^{(2)},\theta_e^{(2)},\theta_i^{(2)},\varphi_R^{(2)}$, $\varphi_e,\varphi_m,\varphi_{ma},\varphi_{mp},\varphi_{ms})$.
We have:
\begin{equation}
\label{eq:xichi}
\begin{aligned}
\rho &=\rho_c+\exp(L_{\xi_{2}})\exp(L_{\xi_{1}})\exp(L_{\chi_{6}})\ldots\exp(L_{\chi_{1}})\delta \rho^{(2)}\\
\varphi&=\exp(L_{\xi_{2}})\exp(L_{\xi_{1}})\exp(L_{\chi_{6}})\ldots\exp(L_{\chi_{1}})\varphi^{(2)}\\
z&=\exp(L_{\xi_{2}})\exp(L_{\xi_{1}})\exp(L_{\chi_{6}})\ldots\exp(L_{\chi_{1}})z^{(2)}\\
p_\rho&=\exp(L_{\xi_{2}})\exp(L_{\xi_{1}})\exp(L_{\chi_{6}})\ldots\exp(L_{\chi_{1}})p_\rho^{(2)}\\
p_\varphi&=p_c+\exp(L_{\xi_{2}})\exp(L_{\xi_{1}})\exp(L_{\chi_{6}})\ldots\exp(L_{\chi_{1}})J_\varphi^{(2)}\\
p_z&=\exp(L_{\xi_{2}})\exp(L_{\xi_{1}})\exp(L_{\chi_{6}})\ldots\exp(L_{\chi_{1}})p_z^{(2)}
\end{aligned}
\end{equation}
where
\begin{equation}
\begin{aligned}
\delta \rho^{(2)}&=\frac{1}{\sqrt{\kappa}}\left[\left(x_{e,f}+\delta x_e^{(2)}\right)\cos(\varphi_R^{(2)}+\varphi_e+\varphi_m)+\left(y_{e,f}+\delta y_e^{(2)}\right)\sin(\varphi_R^{(2)}+\varphi_e+\varphi_m)\right]\\
\varphi^{(2)}&=\varphi_R^{(2)}\\
\delta z^{(2)}&=\frac{1}{\sqrt{\kappa_z}}\left[\left(x_{i,f}+\delta x_i^{(2)}\right)\cos(\varphi_R^{(2)}+\varphi_e)+\left(y_{i,f}+\delta y_i^{(2)}\right)\sin(\varphi_R^{(2)}+\varphi_e)\right]\\
p_\rho^{(2)}&=\sqrt{\kappa}\left[\left(y_{e,f}+\delta y_e^{(2)}\right)\cos(\varphi_R^{(2)}+\varphi_e+\varphi_m)-\left(x_{e,f}+\delta x_e^{(2)}\right)\sin(\varphi_R^{(2)}+\varphi_e+\varphi_m)\right]\\
J_\varphi^{(2)}&=J_R^{(2)}-\frac{1}{2}\left[\left(x_{e,f}+\delta x_e^{(2)}\right)^2+\left(y_{e,f}+\delta y_e^{(2)}\right)^2\right]-\frac{1}{2}\left[\left(x_{i,f}+\delta x_i^{(2)}\right)^2+\left(y_{i,f}+\delta y_i^{(2)}\right)^2\right]\\
p_z^{(2)}&=\sqrt{\kappa_z}\left[\left(y_{i,f}+\delta y_i^{(2)}\right)\cos(\varphi_R^{(2)}+\varphi_e)-\left(x_{i,f}+\delta x_i^{(2)}\right)\sin(\varphi_R^{(2)}+\varphi_e)\right]\\
\end{aligned}
\end{equation}
with
\begin{equation}
\label{eq:bsym2}
\begin{aligned}
\left(
\begin{array}{c}
\delta x_e^{(2)}\\
\delta x_i^{(2)}\\
\delta y_e^{(2)}\\
\delta y_i^{(2)}
\end{array}\right)
=\mathbf{B_{sym}}\left(
\begin{array}{c}
\sqrt{I_e^{(2)}}e^{i\theta_e^{(2)}}\\
\sqrt{I_i^{(2)}}e^{i\theta_i^{(2)}}\\
-i \sqrt{I_e^{(2)}}e^{-i\theta_e^{(2)}}\\
-i \sqrt{I_i^{(2)}}e^{-i\theta_i^{(2)}}
\end{array}\right).
\end{aligned}
\end{equation}
Using Eqs. \eqref{eq:xichi} to \eqref{eq:bsym2}, any state vector expressed in the original variables $(\rho,\varphi,z,p_\rho,p_\varphi,p_z)$ can be expressed in terms of the final variables $(I_e^{(2)},I_i^{(2)},J_R^{(2)},$ $\theta_e^{(2)},\theta_i^{(2)},\varphi_R^{(2)},\varphi_e,\varphi_m,\varphi_{ma},\varphi_{mp},\varphi_{ms})$.\\
Doing the substitution
\begin{equation}
\begin{aligned}
\sqrt{I_e^{(2)}}e^{i\theta_e^{(2)}}& \to q_e^{(2)},\\
-i \sqrt{I_e^{(2)}}e^{-i\theta_e^{(2)}}& \to p_e^{(2)},\\
\sqrt{I_i^{(2)}}e^{i\theta_i^{(2)}}& \to q_i^{(2)},\\
-i \sqrt{I_i^{(2)}}e^{-i\theta_i^{(2)}}& \to p_i^{(2)}\ ,
\end{aligned}
\end{equation}
we can also express the original variables $(\rho,\varphi,z,p_\rho,p_\varphi,p_z)$ in terms of $(q_e^{(2)},q_i^{(2)},p_e^{(2)},p_i^{(2)},J_R^{(2)},\varphi_R^{(2)},$ $\varphi_e,\varphi_m,\varphi_{ma},\varphi_{mp},\varphi_{ms})$.
In particular, if $(q_e^{(2)}(t),q_i^{(2)}(t),p_e^{(2)}(t),p_i^{(2)}(t),J_R^{(2)}(t),\varphi_R^{(2)}(t),\varphi_e(t),\varphi_m(t),$\\$\varphi_{ma}(t),\varphi_{mp}(t),\varphi_{ms}(t))$ is a known solution of the final normal form, it can be expressed, via the above equations, as a time solution  function of the original variables $(\rho(t),\varphi(t),z(t)$, $p_\rho(t),p_\varphi(t),p_z(t))$.

\subsection{Comparison of analytical and numerical results at the forced equilibrium}
\begin{figure}
\begin{minipage}[t]{.48\textwidth}
\includegraphics[width=1\textwidth]{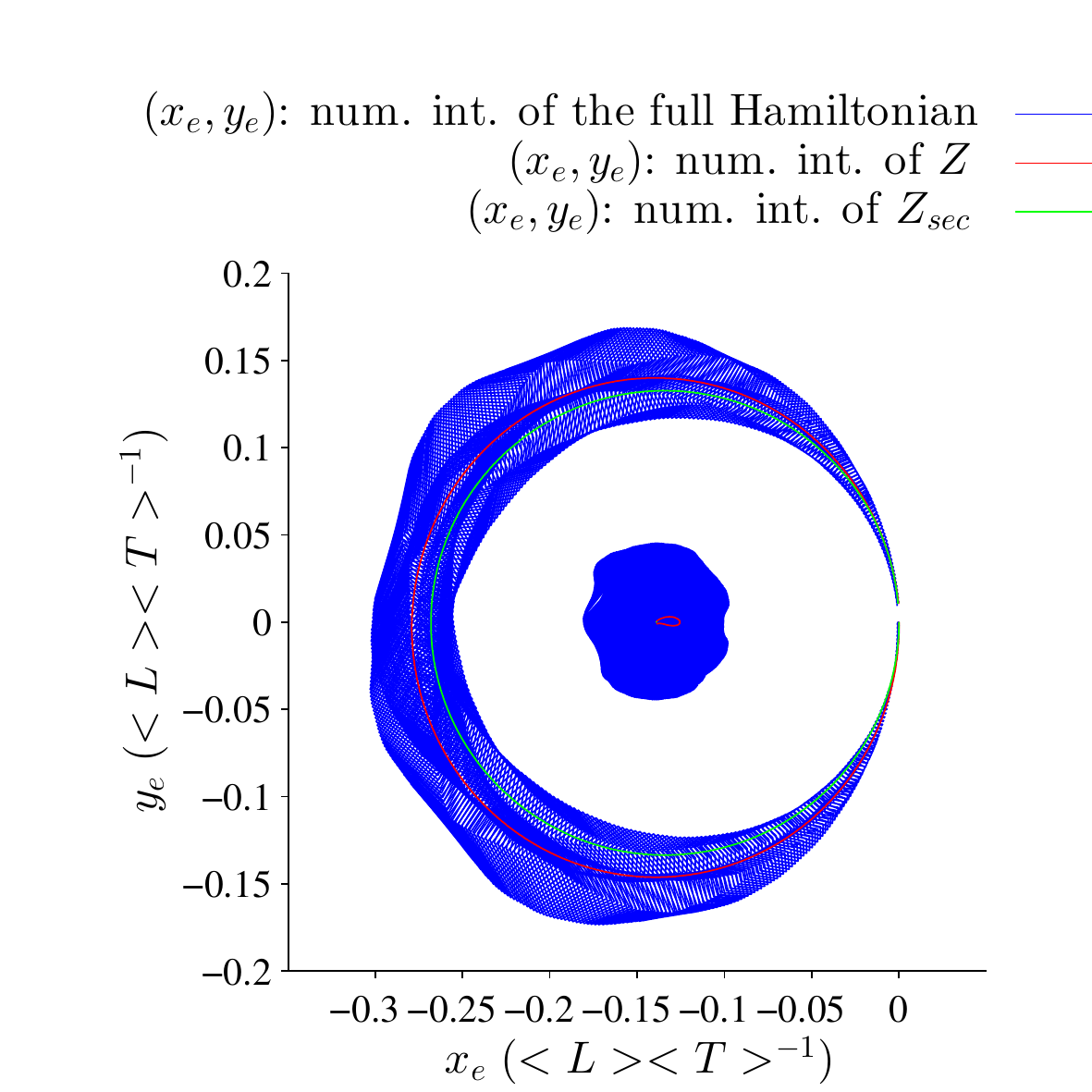}
\end{minipage}
\hfill
\begin{minipage}[t]{.481\textwidth}
\includegraphics[width=1\textwidth]{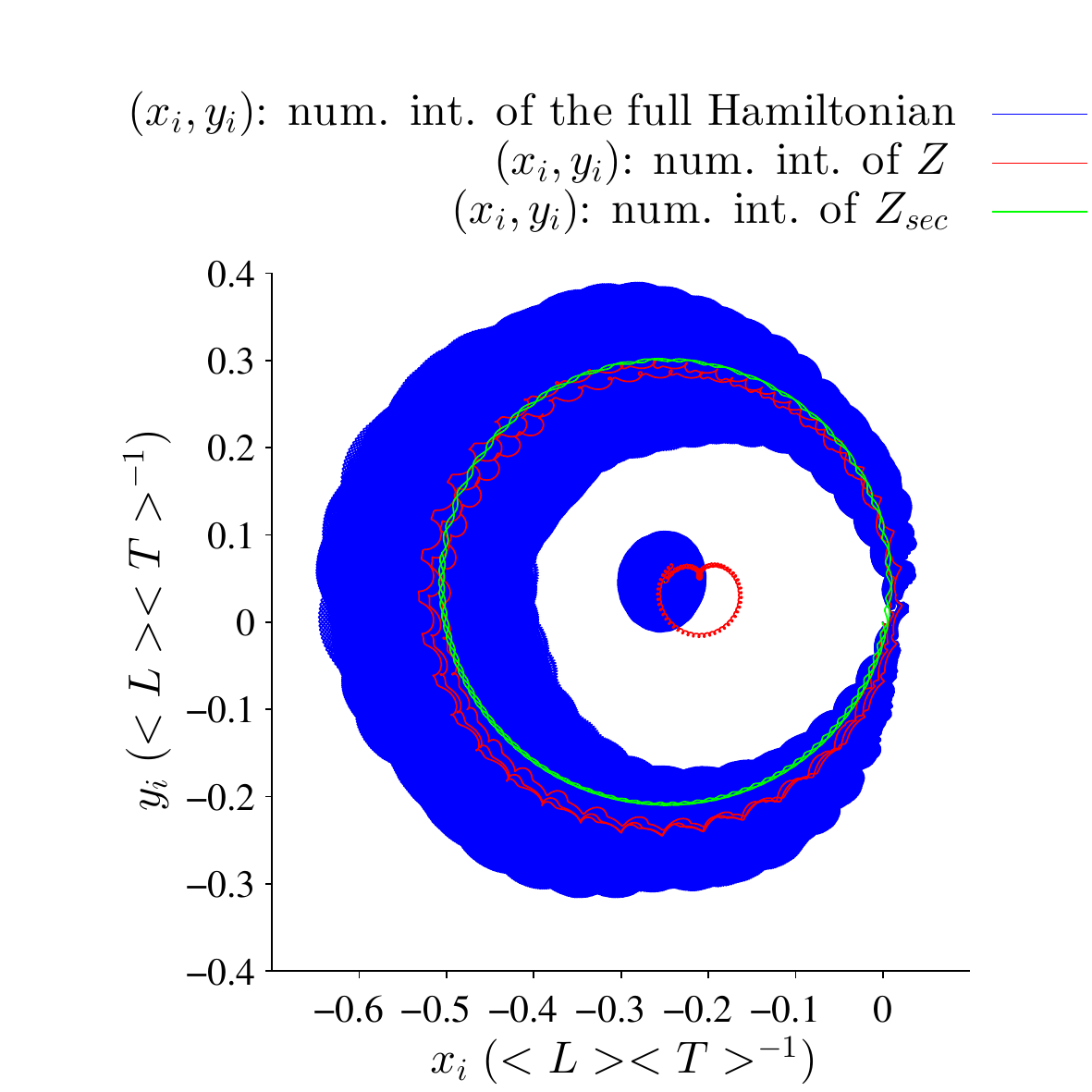}
\end{minipage}
\caption{\label{fig:xeyexiyi10}Representation of two different geostationary orbits with three different methods: Numerical integration of the full Hamiltonian (blue), normal form $Z$ (red), and $Z_{sec}$ (green) for the pair of variables $(x_e,y_e)$ (left, duration $2\pi/\Omega_{e,f} \approx 1$ year) and $(x_i,y_i)$ (right, duration $2\pi/\Omega_{i,f} \approx 38.5$ years).
The outer orbit corresponds to initial conditions $e=0,\, i=0$, the interior one to the forced equilibrium $e = e_{forced}=0.114,\, i = i_{forced}=$ 11.7 deg. The piece of debris considered has an $A/m=10$~m\textsuperscript{2}kg\textsuperscript{-1}.
We notice that for the outer orbit, the normal form $Z$, and $Z_{sec}$ - its component  used to derive the equilibrium - capture the long term dynamics without the daily variations, illustrating the principle of averaging. For the forced equilibrium, we clearly see that it is actually a pseudo equilibrium, since variations do exist for the real (blue) orbit. The motion in $(x_e,y_e)$ and $(x_i,y_i)$ derived from the normal form $Z$ only exhibits tiny variations, since it is an expansion built to find this equilibrium, and the motion derived from $Z_{sec}$ only is actually invisible, since the forced equilibrium is a fixed point for $Z_{sec}$.}
\end{figure}

\begin{figure}
%\captionsetup{justification=centering}
\begin{minipage}[t]{.48\textwidth}
\includegraphics[width=1\textwidth]{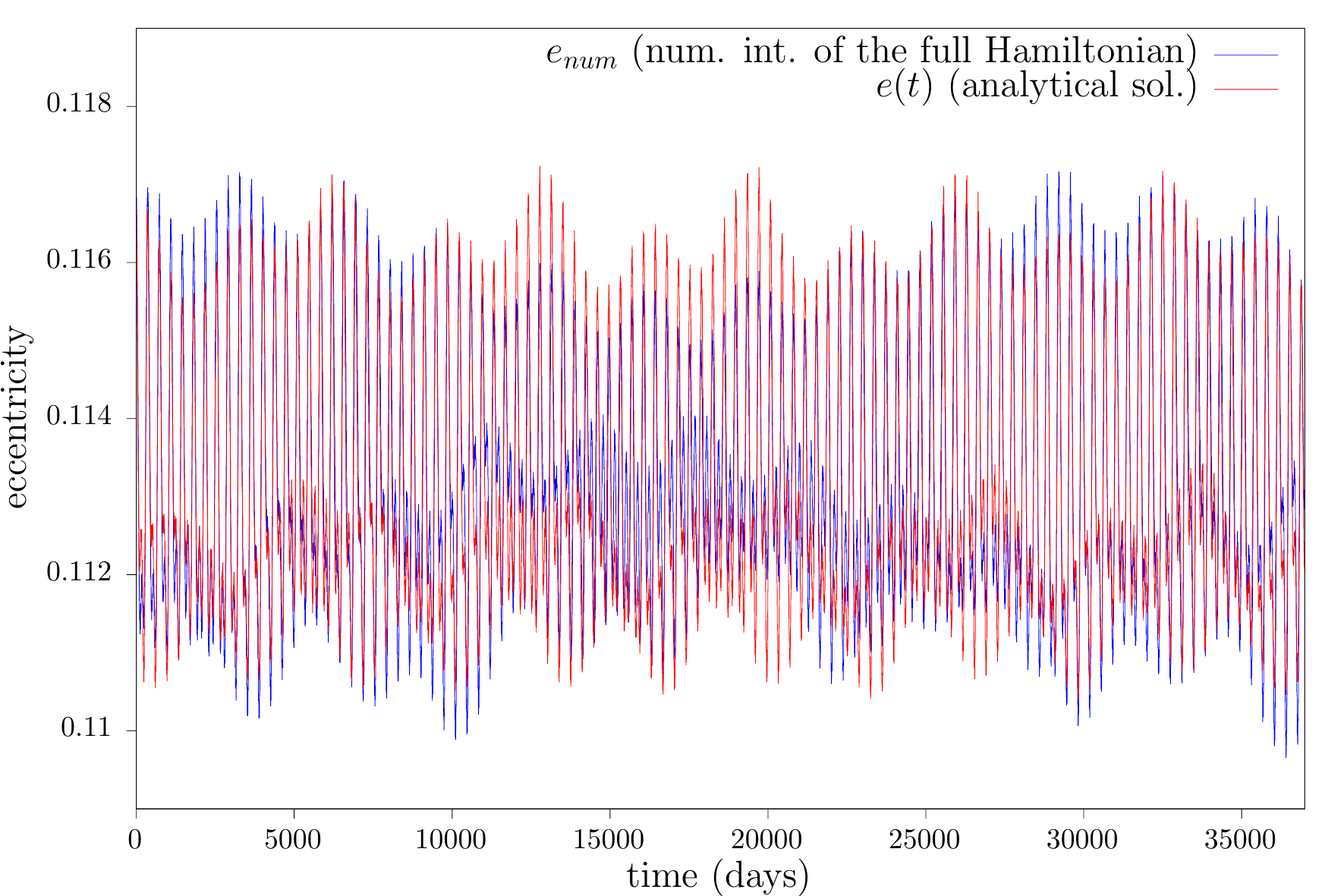}
\end{minipage}
\hfill
\begin{minipage}[t]{.48\textwidth}
\includegraphics[width=1\textwidth]{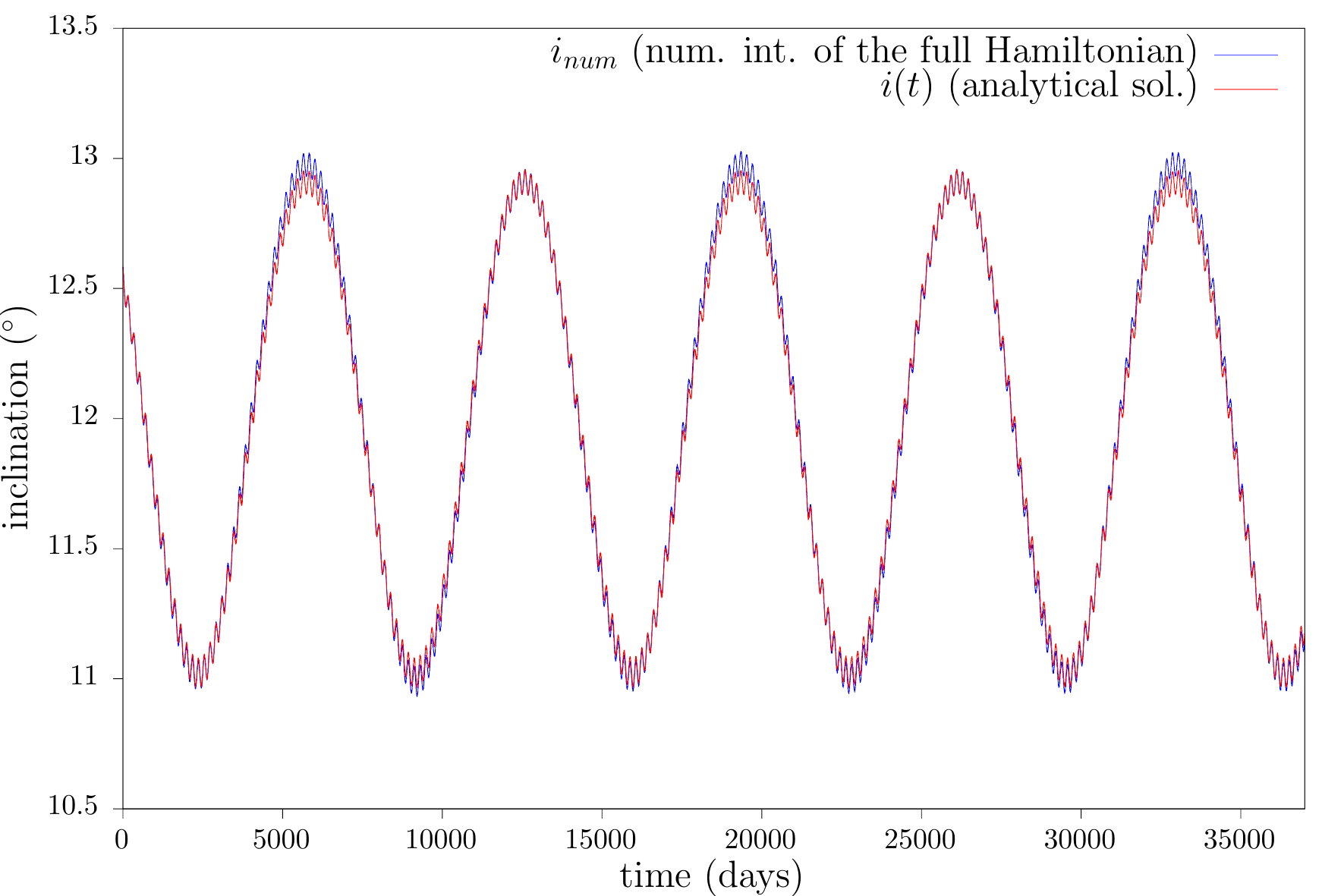}
\end{minipage}
\caption{\label{fig:ei10}Time evolution of the eccentricity (left) and the inclination (right), at the forced equilibrium, showing the numerical integration of the full Hamiltonian (blue) and time-explicit analytical model (red) for an $A/m=10$~m\textsuperscript{2}kg\textsuperscript{-1}. Note the scale on the left plot. The relative error is less than 1.5\% (0.002) for the eccentricity and less than 0.6\% (0.07 deg) for the inclination over 100 years.}
\end{figure}

We now apply these formulas to the forced equilibrium solution.
Implementing the above transformations to the forced solution:
\begin{equation}
\label{eq:finaltransfo}
\begin{aligned}
q_e^{(2)}&=q_i^{(2)}=p_e^{(2)}=p_i^{(2)}=J_R^{(2)}=0\\
\varphi_R^{(2)}&=1.31023 \, \text{rad},\,\varphi_e=\Omega_E t,\,\varphi_m=\Omega_M t,\,\varphi_{m_a}=\Omega_{M_a} t,\,\varphi_{m_p}=\Omega_{M_p} t,\,\varphi_{m_s}=\Omega_{M_s} t
\end{aligned}
\end{equation}
yields the analytical formulas for $(\rho(t),\varphi(t),z(t),p_\rho(t),p_\varphi(t),p_z(t))$ at the forced equilibrium.
Due to \eqref{eq:finaltransfo}, we see that the true nature of the forced equilibrium is a trajectory lying on a 5-dimensional torus, i.e., depending on the five angles evolving with incommensurable frequencies.
Setting $t=0$ we find an analytical  approximation to initial conditions on this special torus solution, which would be hard to recover numerically.
The time evolution of the analytical forced equilibrium solution along with a comparison with the numerical integration of the associated initial conditions under the full Hamiltonian are shown in Figs.~\ref{fig:xeyexiyi10} and \ref{fig:ei10}. We recall here that the analytical expressions have been obtained by taking $N_{pol}=8$, a first normalization of order 6, and a second one of order 2.

First, Figure~\ref{fig:xeyexiyi10} shows the evolution for an object with $A/m=10$~m\textsuperscript{2}kg\textsuperscript{-1}, in the variables $(x_e,y_e)$ on the left and $(x_i,y_i)$ on the right, of two different orbits, using dynamics derived from three different models numerically integrated: i) the full Hamiltonian, ii) the full first normal form $Z$, and iii) $Z_{sec}$.\\
For the left part, showing $(x_e,y_e)$, the propagation time is equal to one year, the period of the forced eccentricity, and for the right part, showing $(x_i,y_i)$, it is about 38.5 years, which is the period of the forced inclination for this $A/m$.
The inner orbits corresponds to a propagation for one year of the initial condition at the forced equilibrium $(x_{e,f},y_{e,f},x_{i,f},y_{i,f})$.
These orbits are clearly quasi-periodic since they exhibit variations in the $(x_e,y_e)$ variables in the full model (blue), while the normal form dynamics $Z$ (in red) shows variations, albeit with a smaller size, by virtue of its definition, and the motion derived from $Z_{sec}$ (green) reduces to a point, the initial condition $(x_{e,f},y_{e,f},x_{i,f},y_{i,f})$ being a true equilibrium for this model.
On the other hand, the outer orbits represent a common satellite orbit starting at $e=0$ and $i=0$. They exhibit daily variations in the case of the full Hamiltonian, but the normal form $Z$ and even $Z_{sec}$ capture the averaged dynamics.

Returning to the analytical torus solution obtained after \eqref{eq:finaltransfo}, the comparison with the numerical integration of the full Hamiltonian model for the variables $\rho(t)$ and $z(t)$ shows that, for an object with $A/m=10$~m\textsuperscript{2}kg\textsuperscript{-1}, the error between the analytical solution and the numerical integration of the full model is less than 1\% (less than 400 km) for $\rho$, and less than 600 km for $z$ over 100 years.

Relatively, Figure~\ref{fig:ei10} shows the long-term evolution of the eccentricity (left) and inclination (right) starting at the forced equilibrium for an object with $A/m=10$~m\textsuperscript{2}kg\textsuperscript{-1}.
The eccentricity varies about 3\% with respect to its initial value over 100 years, and the error between the analytical solution and the numerical integration of the full model is less than 1.5\% (0.0015) over this timespan.
As for the inclination, it varies about $\pm 1$ deg with respect to its initial value over 100 years, and the error between the analytical and numerical solutions is less than 0.6\% (0.07 deg) over this timespan.

The same study has been done for an an object with $A/m=1$~m\textsuperscript{2}kg\textsuperscript{-1}, and the error in eccentricity is less than 3.5\% (0.0004 in this case since the forced eccentricity is $\approx 0.0114$), while the error in inclination is less than 0.25\% (0.03 deg).

\section{Conclusion}
\label{sec:concl}
In the present paper, we applied the methods of canonical perturbation theory in a dynamical problem of timely importance, namely the long-term orbital evolution of satellites or space debris, focusing on the GEO region.
Since dissipation effects are negligible, the orbital motion can be represented as a conservative (Hamiltonian) dynamical system.
Our aim was to develop an analytical theory for the long-term orbital evolution of GEO objects in a realistic model, which includes the Earth's most important multipole potential harmonics, realistic representation of the lunisolar gravitational perturbations and the solar radiation pressure.
Our main results are the following:
We express the Hamiltonian of the system using simple cylindrical coordinates and momenta, easily obtained from standard EME coordinates, thus avoiding the cumbersome algebra involved in transformations from physical coordinates to elements and vice versa.
In view of potential applications of this Hamiltonian, beyond the scope of the present study, we provide its complete expansion in electronic file format as an online supplementary material.
Using this Hamiltonian expansion, we then implemented the method of normal forms obtained via Lie series canonical transformations, in order to arrive at a new, transformed, Hamiltonian (i.e. the `normal form'), which exhibits a secular character, i.e., only pairs of action-angle variables corresponding to `slow' frequencies (of periods beyond one year) appear in it.
Isolating the most important terms of the normal form allows to obtain an even simpler model, which serves as a basis for an analytical theory of the secular dynamics.
The model of Eqs.~\eqref{eq:zsimpledef}-\eqref{eq:zressimple}, in particular, allows to specify a stable equilibrium solution of special interest, called the `forced equilibrium'.
Approximate analytical formulas are provided for this special solution, as well as for `proper elements', i.e., quasi-invariants of the motion around the forced equilibrium, whose distribution of values can be of use in statistical studies of the ensemble evolution of populations of GEO objects like space debris (a study of this form is under way).
Finally, using the inverse of our normalizing transformations, we express the forced equilibrium solution in terms of the original state vector variables.
The analytically computed solution by this process represents an orbit evolving quasi-periodically with five independent frequencies related to the frequencies of the perturbing bodies.
Thus, the forced equilibrium solution lies actually on a 5-dimensional torus embedded in the full 8-dimensional phase space of the complete problem.
We compare our analytical solution with that found by numerical integration of the corresponding orbit under the full Hamiltonian, recovering a precision of about 1\% over a period of 100~yr.
We emphasize the practical aspects of this analysis in issues like the safe end-of-life disposal of artificial GEO objects but also in the investigation of the dynamics of planetary rings \citep{Hedman2009} or in the study of circumplanetary dust \citep{Burns2001}.
In terms of CPU time especially, the evaluation of the analytical solution is almost instantaneous, and the numerical integration of the normal form is orders of magnitude faster than the numerical integration of the full Hamiltonian model.
The accuracy of the analytical solution can also ultimately be improved by going to higher orders in the polynomial approximation made and the normalization order.
As a last note, an important fact to realize is also that the analytical solution is valid in a domain around the forced equilibrium, and therefore in a region around GEO. Future studies could show the use of these solutions for reliable long-term analytical propagation at GEO.

\paragraph*{Acknowledgements}
We would like to thank Aaron Rosengren and Alessandro Rossi for insightful discussions. We also acknowledge an anonymous referee for his/her constructive comments. This work is partially funded by the European Commission FP7, through the Stardust ITN, Grant Agreement 317185.  A.C. and G.P. also acknowledge GNFM/INdAM.

%$$$$$$$$$$$$$$$$$$$$$$$$$$$$$$$$$$$$$$$$$$$$$$$$$$$$$$$$$$$$$$$$$$$$$$$$$$$$$$$$$$$$$$$$
\bibliographystyle{plainnat}
\bibliography{paper_eqHAMRGEO_bibli}

\begin{thebibliography}{36}
\providecommand{\natexlab}[1]{#1}
\providecommand{\url}[1]{\texttt{#1}}
\expandafter\ifx\csname urlstyle\endcsname\relax
  \providecommand{\doi}[1]{doi: #1}\else
  \providecommand{\doi}{doi: \begingroup \urlstyle{rm}\Url}\fi

\bibitem[Allan and Cook(1964)]{AllanCook1964}
R.~R. Allan and G.~E. Cook.
\newblock {The Long-Period Motion of the Plane of a Distant Circular Orbit}.
\newblock \emph{Proceedings of the Royal Society A: Mathematical, Physical and
  Engineering Sciences}, 280\penalty0 (1380):\penalty0 97--109, jul 1964.
\newblock ISSN 1364-5021.
\newblock \doi{10.1098/rspa.1964.0133}.
\newblock URL
  \url{http://rspa.royalsocietypublishing.org/cgi/doi/10.1098/rspa.1964.0133}.

\bibitem[Allan and Cook(1967)]{AllanCook1967}
R.~R. Allan and G.~E. Cook.
\newblock {Discussion of paper by S. J. Peale, ‘Dust belt of the Earth'}.
\newblock \emph{Journal of Geophysical Research}, 72\penalty0 (3):\penalty0
  1124--1127, feb 1967.
\newblock ISSN 01480227.
\newblock \doi{10.1029/JZ072i003p01124}.
\newblock URL \url{http://doi.wiley.com/10.1029/JZ072i003p01124}.

\bibitem[Anselmo and Pardini(2005)]{Anselmo2005}
L.~Anselmo and C.~Pardini.
\newblock {Orbital Evolution of Geosynchronous Objects with High Area-To-Mass
  Ratios}.
\newblock In D.~Danesy, editor, \emph{4th European Conference on Space Debris},
  volume 587 of \emph{ESA Special Publication}, page 279, August 2005.

\bibitem[Anselmo and Pardini(2008)]{Anselmo2008}
L.~Anselmo and C.~Pardini.
\newblock {Space debris mitigation in geosynchronous orbit}.
\newblock \emph{Advances in Space Research}, 41\penalty0 (7):\penalty0
  1091--1099, 2008.
\newblock ISSN 02731177.
\newblock \doi{10.1016/j.asr.2006.12.018}.
\newblock URL
  \url{http://linkinghub.elsevier.com/retrieve/pii/S0273117706007757}.

\bibitem[Burns et~al.(2001)Burns, Hamilton, and Showalter]{Burns2001}
JA~Burns, DP~Hamilton, and MR~Showalter.
\newblock {Dusty rings and circumplanetary dust: Observations and simple
  physics}.
\newblock \emph{Interplanetary Dust}, pages 1--85, 2001.
\newblock URL
  \url{http://link.springer.com/chapter/10.1007/978-3-642-56428-4{\_}13}.

\bibitem[Casanova et~al.(2015)Casanova, Petit, and
  Lema{\^{\i}}tre]{Casanova2014}
D.~Casanova, A.~Petit, and A.~Lema{\^{\i}}tre.
\newblock {Long-term evolution of space debris under the J2 effect, the solar
  radiation pressure and the solar and lunar perturbations}.
\newblock \emph{Celestial Mechanics and Dynamical Astronomy}, 123\penalty0
  (2):\penalty0 223--238, October 2015.
\newblock ISSN 0923-2958.
\newblock \doi{10.1007/s10569-015-9644-1}.
\newblock URL \url{"http://dx.doi.org/10.1007/s10569-015-9644-1
  http://link.springer.com/10.1007/s10569-015-9644-1}.

\bibitem[Celletti and Gale\cb{s}(2014)]{Celletti2014}
A.~Celletti and C.~Gale\cb{s}.
\newblock {On the Dynamics of Space Debris: 1:1 and 2:1 Resonances}.
\newblock \emph{Journal of Nonlinear Science}, 24\penalty0 (6):\penalty0
  1231--1262, 2014.
\newblock ISSN 0938-8974.
\newblock \doi{10.1007/s00332-014-9217-6}.
\newblock URL \url{http://link.springer.com/10.1007/s00332-014-9217-6}.

\bibitem[Chao(2005)]{Chao2005}
C.C. Chao.
\newblock \emph{{Applied Orbit Perturbation and Maintenance}}.
\newblock AIAA, November 2005.
\newblock ISBN 978-1-884989-17-9.
\newblock \doi{10.2514/4.989179}.

\bibitem[Chao(2006)]{Chao2006}
C.C. Chao.
\newblock {Analytical Investigation of GEO Debris with High Area-to-Mass
  Ratio}.
\newblock In \emph{AIAA/AAS Astrodynamics Specialist Conference and Exhibit},
  pages 1--9, Reston, Virginia, August 2006. AIAA.
\newblock ISBN 978-1-62410-048-2.
\newblock \doi{10.2514/6.2006-6514}.

\bibitem[Chao and Baker(1983)]{ChaoBaker}
C.C. Chao and J.M. Baker.
\newblock {On the propagation and control of geosynchronous orbits}.
\newblock \emph{Journal of the Astronautical Sciences}, 31:\penalty0 99--115,
  March 1983.

\bibitem[Deprit(1969)]{Deprit1969}
A.~Deprit.
\newblock {Canonical transformations depending on a small parameter}.
\newblock \emph{Celestial Mechanics}, 1\penalty0 (1):\penalty0 12--30, March
  1969.
\newblock ISSN 0008-8714.
\newblock \doi{10.1007/BF01230629}.
\newblock URL \url{http://link.springer.com/10.1007/BF01230629}.

\bibitem[Efthymiopoulos(2012)]{efthyLaPlata}
C.~Efthymiopoulos.
\newblock {Canonical perturbation theory; stability and diffusion in
  Hamiltonian systems: applications in dynamical astronomy}.
\newblock \emph{in Cincotta P. M., Giordano C.M., Efthymiopoulos C., eds, Third
  La Plata International School on Astronomy and Geophysics, Workshop Series of
  the Asociacion Argentina de Astronomia}, 3:\penalty0 3--146, 2012.

\bibitem[Friesen et~al.(1992)]{Friesen1992}
L.J. Friesen et~al.
\newblock {Results in orbital evolution of objects in the geosynchronous
  region}.
\newblock \emph{Journal of Guidance, Control, and Dynamics}, 15\penalty0
  (1):\penalty0 263--267, jan 1992.
\newblock ISSN 0731-5090.
\newblock \doi{10.2514/3.20827}.
\newblock URL \url{http://arc.aiaa.org/doi/abs/10.2514/3.20827}.

\bibitem[Giorgilli(2002)]{giorgilli2002}
A.~Giorgilli.
\newblock Notes on exponential stability of hamiltonian systems.
\newblock \emph{Lectures presented at the Research Trimester in Dynamical
  systems, 'Centro di Ricerca matematica Ennio De Giorgi', Pisa}, 2002.

\bibitem[Hedman et~al.(2009)Hedman, Burns, Tiscareno, and Porco]{Hedman2009}
M.M. Hedman, J.A. Burns, M.S. Tiscareno, and C.C. Porco.
\newblock {Organizing some very tenuous things: Resonant structures in Saturn's
  faint rings}.
\newblock \emph{Icarus}, 202\penalty0 (1):\penalty0 260--279, jul 2009.
\newblock ISSN 00191035.
\newblock \doi{10.1016/j.icarus.2009.02.016}.
\newblock URL \url{http://dx.doi.org/10.1016/j.icarus.2009.02.016
  http://linkinghub.elsevier.com/retrieve/pii/S0019103509000785}.

\bibitem[Hori(1966)]{Hori1966}
G.-I. Hori.
\newblock {Theory of general perturbation with unspecified canonical variable}.
\newblock \emph{Publications of the Astronomical Society of Japan}, 18\penalty0
  (4):\penalty0 287--296, 1966.
\newblock ISSN 0004-6264.
\newblock URL \url{http://adsabs.harvard.edu/full/1966PASJ...18..287H7}.

\bibitem[Hubaux and Lema{\^{\i}}tre(2013)]{Hubaux2013}
C.~Hubaux and A.~Lema{\^{\i}}tre.
\newblock {The impact of Earth’s shadow on the long-term evolution of space
  debris}.
\newblock \emph{Celestial Mechanics and Dynamical Astronomy}, 116\penalty0
  (1):\penalty0 79--95, May 2013.
\newblock ISSN 0923-2958.
\newblock \doi{10.1007/s10569-013-9480-0}.
\newblock URL \url{http://link.springer.com/10.1007/s10569-013-9480-0}.

\bibitem[{Kaula}(1966)]{Kaula1966}
W.~M. {Kaula}.
\newblock \emph{{Theory of satellite geodesy. Applications of satellites to
  geodesy}}.
\newblock Waltham, Mass.: Blaisdell, 1966.

\bibitem[{Kubo-Oka} and {Sengoku}(1999)]{Kubo1999}
T.~{Kubo-Oka} and A.~{Sengoku}.
\newblock {Solar radiation pressure model for the relay satellite of SELENE}.
\newblock \emph{Earth, Planets, and Space}, 51:\penalty0 979--986, 1999.

\bibitem[Lemoine et~al.(1998)]{EGM96}
F.G. Lemoine et~al.
\newblock {The development of the joint NASA GSFC and the National Imagery and
  Mapping Agency (NIMA) geopotential model EGM96}.
\newblock \emph{NASA Tech. Publ., TP-1998-206861, NASA Goddard Space Flight
  Cent., Washington, D. C.}, 1998.

\bibitem[Liou and Weaver(2005)]{Liou2005}
J.-C. Liou and J.K. Weaver.
\newblock {Orbital Dynamics of High Area-To-Mass Ratio Debris and Their
  Distribution in the Geosynchronous Region}.
\newblock In D.~Danesy, editor, \emph{4th European Conference on Space Debris},
  volume 587 of \emph{ESA Special Publication}, page 285, August 2005.

\bibitem[McMahon and Scheeres(2010)]{McMahon2010}
J.W. McMahon and D.J. Scheeres.
\newblock {Secular orbit variation due to solar radiation effects: A detailed
  model for BYORP}.
\newblock \emph{Celestial Mechanics and Dynamical Astronomy}, 106\penalty0
  (3):\penalty0 261--300, 2010.
\newblock ISSN 09232958.
\newblock \doi{10.1007/s10569-009-9247-9}.

\bibitem[Mignard and Henon(1984)]{MignardHenon1984}
F.~Mignard and M.~Henon.
\newblock {About an unsuspected integrable problem}.
\newblock \emph{Celestial Mechanics}, 33\penalty0 (3):\penalty0 239--250, jul
  1984.
\newblock ISSN 0008-8714.
\newblock \doi{10.1007/BF01230506}.
\newblock URL \url{http://link.springer.com/10.1007/BF01230506}.

\bibitem[Montenbruck and Gill(2000)]{Montenbruck2000}
O.~Montenbruck and E.~Gill.
\newblock \emph{{Satellite Orbits}}, volume 134.
\newblock Springer Berlin Heidelberg, July 2000.
\newblock ISBN 978-3-642-63547-2.
\newblock \doi{10.1007/978-3-642-58351-3}.

\bibitem[Murray and Dermott(1999)]{MurrayDermott1999}
C.D. Murray and S.F. Dermott.
\newblock \emph{{Solar System Dynamics}}.
\newblock Springer Berlin Heidelberg, 1999.
\newblock ISBN 0 521 57597.

\bibitem[Musen(1961)]{Musen1961b}
P.~Musen.
\newblock {On the long-period lunar and solar effects on the motion of an
  artificial satellite: 2.}
\newblock \emph{Journal of Geophysical Research}, 66\penalty0 (9):\penalty0
  2797--2805, sep 1961.
\newblock ISSN 01480227.
\newblock \doi{10.1029/JZ066i009p02797}.
\newblock URL \url{http://doi.wiley.com/10.1029/JZ066i009p02797}.

\bibitem[Roncoli(2005)]{Roncoli2005}
R.B. Roncoli.
\newblock {Lunar constants and models document}.
\newblock \emph{JPL D-32296, Jet Propulsion Laboratory, Pasadena, CA}, 2005.

\bibitem[Rosengren and Scheeres(2013)]{Rosengren2013}
A.J. Rosengren and D.J. Scheeres.
\newblock {Long-term dynamics of high area-to-mass ratio objects in high-Earth
  orbit}.
\newblock \emph{Advances in Space Research}, 52\penalty0 (8):\penalty0
  1545--1560, October 2013.
\newblock ISSN 02731177.
\newblock \doi{10.1016/j.asr.2013.07.033}.
\newblock URL
  \url{http://linkinghub.elsevier.com/retrieve/pii/S0273117713004626}.

\bibitem[Rosengren et~al.(2014)Rosengren, Scheeres, and
  McMahon]{Rosengren2014b}
A.J. Rosengren, D.J. Scheeres, and J.W. McMahon.
\newblock {The classical Laplace plane as a stable disposal orbit for
  geostationary satellites}.
\newblock \emph{Advances in Space Research}, 53\penalty0 (8):\penalty0
  1219--1228, apr 2014.
\newblock ISSN 02731177.
\newblock \doi{10.1016/j.asr.2014.01.034}.
\newblock URL
  \url{http://linkinghub.elsevier.com/retrieve/pii/S027311771400088X}.

\bibitem[Rosengren et~al.(2015)Rosengren, Alessi, Rossi, and
  Valsecchi]{Rosengren2015a}
A.J. Rosengren, E.M. Alessi, A.~Rossi, and G.B. Valsecchi.
\newblock {Chaos in navigation satellite orbits caused by the perturbed motion
  of the Moon}.
\newblock \emph{Monthly Notices of the Royal Astronomical Society},
  449\penalty0 (4):\penalty0 3522--3526, 2015.
\newblock ISSN 0035-8711.
\newblock \doi{10.1093/mnras/stv534}.
\newblock URL
  \url{http://arxiv.org/abs/1503.02581$\backslash$nhttp://dx.doi.org/10.1093/mnras/stv534}.

\bibitem[Schildknecht(2007)]{Schildknecht2007}
T.~Schildknecht.
\newblock {Optical surveys for space debris}.
\newblock \emph{The Astronomy and Astrophysics Review}, 14\penalty0
  (1):\penalty0 41--111, jan 2007.
\newblock ISSN 0935-4956.
\newblock \doi{10.1007/s00159-006-0003-9}.
\newblock URL \url{http://link.springer.com/10.1007/s00159-006-0003-9}.

\bibitem[Schildknecht et~al.(2004)]{Schildknecht2004}
T.~Schildknecht et~al.
\newblock {Optical observations of space debris in GEO and in highly-eccentric
  orbits}.
\newblock \emph{Advances in Space Research}, 34\penalty0 (5):\penalty0
  901--911, jan 2004.
\newblock ISSN 02731177.
\newblock \doi{10.1016/j.asr.2003.01.009}.
\newblock URL
  \url{http://linkinghub.elsevier.com/retrieve/pii/S0273117704000651}.

\bibitem[Tremaine et~al.(2009)Tremaine, Touma, and Namouni]{Tremaine2009}
S.~Tremaine, J.~Touma, and F.~Namouni.
\newblock {SATELLITE DYNAMICS ON THE LAPLACE SURFACE}.
\newblock \emph{The Astronomical Journal}, 137\penalty0 (3):\penalty0
  3706--3717, mar 2009.
\newblock ISSN 0004-6256.
\newblock \doi{10.1088/0004-6256/137/3/3706}.
\newblock URL
  \url{http://stacks.iop.org/1538-3881/137/i=3/a=3706?key=crossref.7166da0a6724d0175f85d1c362264064}.

\bibitem[Valk et~al.(2008)Valk, Lema{\^{i}}tre, and Anselmo]{Valk2008a}
S.~Valk, A.~Lema{\^{i}}tre, and L.~Anselmo.
\newblock {Analytical and semi-analytical investigations of geosynchronous
  space debris with high area-to-mass ratios}.
\newblock \emph{Advances in Space Research}, 41\penalty0 (7):\penalty0
  1077--1090, jan 2008.
\newblock ISSN 02731177.
\newblock \doi{10.1016/j.asr.2007.10.025}.
\newblock URL
  \url{http://linkinghub.elsevier.com/retrieve/pii/S0273117707010435}.

\bibitem[Valk et~al.(2009{\natexlab{a}})Valk, Delsate, Lema{\^{i}}tre, and
  Carletti]{Valk2009c}
S.~Valk, N.~Delsate, A.~Lema{\^{i}}tre, and T.~Carletti.
\newblock {Global dynamics of high area-to-mass ratios GEO space debris by
  means of the MEGNO indicator}.
\newblock \emph{Advances in Space Research}, 43\penalty0 (10):\penalty0
  1509--1526, 2009{\natexlab{a}}.
\newblock ISSN 02731177.
\newblock \doi{10.1016/j.asr.2009.02.014}.
\newblock URL
  \url{http://www.sciencedirect.com/science/article/pii/S0273117709001471}.

\bibitem[Valk et~al.(2009{\natexlab{b}})Valk, Lema{\^{i}}tre, and
  Deleflie]{Valk2009d}
S.~Valk, A.~Lema{\^{i}}tre, and F.~Deleflie.
\newblock {Semi-analytical theory of mean orbital motion for geosynchronous
  space debris under gravitational influence}.
\newblock \emph{Advances in Space Research}, 43\penalty0 (7):\penalty0
  1070--1082, 2009{\natexlab{b}}.
\newblock ISSN 02731177.
\newblock \doi{10.1016/j.asr.2008.12.015}.
\newblock URL
  \url{http://linkinghub.elsevier.com/retrieve/pii/S0273117708006807}.

\end{thebibliography}
%$$$$$$$$$$$$$$$$$$$$$$$$$$$$$$$$$$$$$$$$$$$$$$$$$$$$$$$$$$$$$$$$$$$$$$$$$$$$$$$$$$$$$$$$

\clearpage

%\appendix
\section*{Appendix A: Book-keeping for the solar and lunar position vectors}
In this section we give the position vectors of the Sun and the Moon along with their appropriate book-keeping orders.\\
We have
\begin{equation}
\label{eq:rsunvec}
\mathbf{r_\odot}=
R_{G} \cdot \left(
\begin{array}{c}
r_\odot \cos \lambda_\odot\\
r_\odot \sin \lambda_\odot (\lambda_{ls}^2 (\cos \varepsilon-1)+1)\\
r_\odot \sin \lambda_\odot \lambda_{ls} \sin \varepsilon\
\end{array} \right),
\end{equation}
where $\varepsilon=23.43929111^\circ$ is the obliquity of the ecliptic, and $\lambda_\odot$,  $r_\odot$ are the longitude and radial distance of the Sun in the EME2000 frame. The time evolution of $\lambda_\odot$ and $r_\odot$ is given by the following truncated series expansions (\citet{Montenbruck2000}, p. 71):
\begin{equation}
\label{eq:rlambdasun}
\begin{aligned}
\lambda_\odot &=\Omega+\omega+M+\lambda_{ls}6892'' \sin M ++\lambda_{ls}^2 72'' \sin 2M\\
r_\odot &= 149.619-2.499\lambda_{ls} \cos M - 0.021 \lambda_{ls}^2 \cos 2M
\end{aligned}
\end{equation}
with $\Omega_\odot$ the longitude of the ascending node, $\omega_\odot$ the argument of periapse and $M_{\odot}$ the mean anomaly of the Sun. Their values are:
\begin{equation}
\label{eq:Sunelts}
\Omega_\odot+\omega_\odot = 282.9400^\circ \ , \quad 
M_{\odot} = \varphi_M +357.5256^\circ \ ,
\end{equation}
with
\begin{equation}
\varphi_M=\Omega_M t \ ,
\end{equation}
and 
\begin{equation}
\label{eq:ratesun}
\Omega_M=359.99049^\circ \ \text{yr}^{-1}\ ,
\end{equation}
the yearly frequency with which the Sun revolves around the Earth in the geocentric frame.
Finally, these formulas represent the Sun moving on a fixed ellipse of eccentricity  $e_\odot=0.016709$.\\

The Moon's motion with respect to the Earth is more complex. We have:
\begin{equation}
\label{eq:rMoonvec}
\mathbf{r_{\leftmoon}}=
R_{G} \cdot \left(
\begin{array}{ccc}
1 & 0 & 0 \\
0 & \lambda_{ls}^2 (\cos \varepsilon-1)+1 & -\lambda_{ls} \sin\varepsilon\\
0 & \lambda_{ls} \sin\varepsilon & \lambda_{ls}^2(\cos \varepsilon-1)+1
\end{array} \right)
\
\left(
\begin{array}{c}
r_{\leftmoon} \cos \lambda_{\leftmoon} \cos \beta_{\leftmoon} \\
r_{\leftmoon} \sin \lambda_{\leftmoon} \cos \beta_{\leftmoon}\\
r_{\leftmoon} \sin \beta_{\leftmoon}
\end{array} \right)\ ,
\end{equation}
where $r_{\leftmoon}$ is the Moon's distance, and $\lambda_{\leftmoon}$, $\beta_{\leftmoon}$ are the Moon's longitude and latitude, both in the EME2000 frame. The time evolution of these quantities is represented by the following truncated series expansions (\citet{Montenbruck2000} p. 72):
\begin{equation}
\label{eq:rmoon}
\begin{aligned}
r_{\leftmoon} &= (385000-20905 \lambda_{ls} \cos(l_{{\leftmoon}})-3699 \lambda_{ls}^2 \cos(2 D_{{\leftmoon}}-l_{{\leftmoon}})\\
&-2956 \lambda_{ls}^2 \cos(2 D_{{\leftmoon}})-570 \lambda_{ls}^2 \cos(2 l_{{\leftmoon}})\\
&+246 \lambda_{ls}^2 \cos(2 l_{{\leftmoon}}-2 D_{{\leftmoon}})-205 \lambda_{ls}^2 \cos(l'_{{\leftmoon}}-2 D_{{\leftmoon}})\\
&-171 \lambda_{ls}^2 \cos(l_{{\leftmoon}}+2 D_{{\leftmoon}})\\
&-152 \lambda_{ls}^2 \cos(l_{{\leftmoon}}+l'_{{\leftmoon}}-2 D_{{\leftmoon}})) \ \text{km}\\
\end{aligned}
\end{equation}
\begin{equation}
\label{eq:lambdamoon}
\begin{aligned}
\lambda_{\leftmoon} &= L_0+ 22640'' \lambda_{ls} \sin(l_{\leftmoon}) + 769'' \lambda_{ls}^2 \sin(2 l_{\leftmoon})\\
& - 4856'' \lambda_{ls}^2 \sin(l_{\leftmoon} - 2 D_{\leftmoon}) + 2370'' \lambda_{ls}^2 \sin(2 D_{\leftmoon})\\
& - 668'' \lambda_{ls}^2 \sin(l'_{\leftmoon}) - 412'' \lambda_{ls}^2 \sin(2 F_{\leftmoon})\\
& - 212'' \lambda_{ls}^2 \sin(2 l_{\leftmoon} - 2 D_{\leftmoon}) - 206'' \lambda_{ls}^2 \sin(l_{\leftmoon} + l'_{\leftmoon} - 2 D_{\leftmoon})\\
& + 192'' \lambda_{ls}^2 \sin(l_{\leftmoon} + 2 D_{\leftmoon}) - 165'' \lambda_{ls}^2 \sin(l'_{\leftmoon} - 2 D_{\leftmoon})\\
& + 148'' \lambda_{ls}^2 \sin(l_{\leftmoon} - l'_{\leftmoon}) - 125'' \lambda_{ls}^2 \sin(D_{\leftmoon})\\
& - 110'' \lambda_{ls}^2 \sin(l_{\leftmoon} + l'_{\leftmoon}) - 55'' \lambda_{ls}^2 \sin(2 F_{\leftmoon} - 2 D_{\leftmoon})\\
\end{aligned}
\end{equation}
\begin{equation}
\label{eq:betamoon}
\begin{aligned}
\beta_{\leftmoon} &= \lambda_{ls} 18520'' \sin(F_{\leftmoon} + \lambda_{\leftmoon}-L_0 + 412'' \lambda_{ls} \sin(2F_{\leftmoon})\\
& + 541'' \lambda_{ls} \sin(l'_{\leftmoon})) - 526'' \lambda_{ls}^2 \sin(F_{\leftmoon} - 2D_{\leftmoon})\\
& + 44'' \lambda_{ls}^2 \sin(l_{\leftmoon} + F_{\leftmoon} - 2D_{\leftmoon}) - 31'' \lambda_{ls}^2 \sin(-l_{\leftmoon} + F_{\leftmoon} - 2D_{\leftmoon})\\
&- 25'' \lambda_{ls}^2 \sin(-2l_{\leftmoon} + F_{\leftmoon})- 23'' \lambda_{ls}^2 \sin(l'_{\leftmoon} + F_{\leftmoon} - 2D_{\leftmoon})\\
&+ 21'' \lambda_{ls}^2 \sin(-l_{\leftmoon} + F_{\leftmoon}) + 11'' \lambda_{ls}^2 \sin(-l'_{\leftmoon} + F_{\leftmoon} - 2D_{\leftmoon})\ .
\end{aligned}
\end{equation}
where $L_0$ is the Moon's mean longitude, $l_{\leftmoon}$ is the Moon's mean anomaly, $l'_{\leftmoon}$ is equal to the Sun's mean anomaly, $F_{\leftmoon}$ is the the mean angular distance of the Moon from the ascending node and $D_{\leftmoon}$ is the difference between the mean longitudes of the Sun and the Moon. We also define for future reference $a_{\leftmoon}=385000 \, \text{km}$ the mean distance of the Moon, and $e_{\leftmoon}=0.055$ its eccentricity.
We have:
\begin{equation}\label{eq:moonphases}
\begin{aligned}
L_0&=\varphi_{M_p}+\varphi_{M_a}+218.31617^\circ\\
l_{\leftmoon}&=\varphi_{M_a}+134.96292^\circ\\
l'_{\leftmoon}&=M_{\odot}=\varphi_M+357.52543^\circ\\
F_{\leftmoon}&=\varphi_{M_p}+\varphi_{M_a}+\varphi_{M_s}+93.27283^\circ\\
D_{\leftmoon}&=\varphi_{M_p}+\varphi_{M_a}-\varphi_{M}+297.85027^\circ
\end{aligned}
\end{equation}
with
\begin{equation}
\begin{aligned}
\varphi_{M_a}&=\Omega_{M_a} \, t\\
\varphi_{M_p}&=\Omega_{M_p} \, t\\
\varphi_{M_s}&=\Omega_{M_s} \, t\\
\end{aligned}
\end{equation}
and
\begin{equation}
\label{eq:rates}
\begin{aligned}
\Omega_{M_a}&=4771.9886753^\circ \ \text{yr}^{-1}\\
\Omega_{M_p}&=40.6901335^\circ \ \text{yr}^{-1}\\
\Omega_{M_s}&=19.3413784^\circ \ \text{yr}^{-1}.
\end{aligned}
\end{equation}

\paragraph{Units}
Given the various unit systems used in the literature for obtaining the above expressions, Table \ref{tab:units} summarizes the constants appearing in the previous subsections in a convenient unified system of units.
We set the unit of time $<T>$ equal to one synodic Earth day (equal to 86400~s, hereafter referred to as `day'), and the unit of the velocity equal to $<V>=1$~km~s\textsuperscript{-1}. This sets the length unit $<L>=86400$~km.
Finally, all model's angular variables are measured in radians, and all angular variables appearing in Eqs.~\eqref{eq:rsunvec} to \eqref{eq:moonphases} are transformed to radians.

\begin{table}[h!]
\centering
\caption{Values of the constants in the unified system of units}
\begin{tabular}{ll}
\hline\hline
$\mu_\oplus$ & $4.613431039 <L>^3<T>^{-2}$ \\
$R_\oplus$ & $0.073821 <L>$ \\
$\mu_\odot$ & $1536023.61132 <L>^3<T>^{-2}$ \\
$\mu_{\leftmoon}$ & $0.0567454  <L>^3<T>^{-2}$ \\
$\Omega_E$ & $6.300388~\mbox{rad}<T>^{-1}$ \\
$\Omega_M$ & $0.0172019~\mbox{rad}<T>^{-1}$ \\
$\Omega_{M_a}$ & $0.2280271437~\mbox{rad}<T>^{-1}$ \\
$\Omega_{M_p}$ & $0.00194435~\mbox{rad}<T>^{-1}$ \\
$\Omega_{M_s}$ & $0.00092421~\mbox{rad}<T>^{-1}$ \\
$P_r$ & $3.93 \times 10^{-4}~(kg)<L><T>^{-2}$ \\
$a_\odot$ & $1731.70<L>$ \\
$a_{\leftmoon}$ & $4.45602<L>$ \\
\hline\hline
\end{tabular}
\label{tab:units}
\end{table}%\vfill

\section*{Appendix B: Canonical normalization via Lie series}
In this section we explicit the process of canonical normalization via Lie series.\\
In this method, we seek to find a sequence of consecutive near-identity canonical transformations
\begin{equation}
\label{eq:translie}
(\boldsymbol{\varphi},\boldsymbol{J})
\equiv (\boldsymbol{\varphi}^{(0)},\boldsymbol{J}^{(0)})
\rightarrow (\boldsymbol{\varphi}^{(1)},\boldsymbol{J}^{(1)})
\rightarrow (\boldsymbol{\varphi}^{(2)},\boldsymbol{J}^{(2)})
\rightarrow\ldots
\end{equation}
such that, after $r$ steps, the Hamiltonian, transformed in the new variables, takes the form
\begin{equation}
\label{eq:hamnf}
H^{(r)}=Z_0+\lambda Z_{1}+\ldots+\lambda^{r} Z_{r} + \lambda^{r+1} H^{(r)}_{r+1} + \lambda^{r+2} H^{(r)}_{r+2}+\ldots\ 
\end{equation}
The sequence of transformations \eqref{eq:translie} is defined in such a way that the quantity
\begin{equation}
\label{eq:nfm}
Z^{(r)}=Z_0+\lambda Z_{1}+\ldots+\lambda^{r} Z_{r}
\end{equation}
called the `normal form', represents a Hamiltonian function whose dynamics is simpler to analyze than in the original Hamiltonian model. Back-transforming to the original variables, this allows to obtain an approximation of the dynamics of the original system as well. The difference between the true dynamics and the one induced by the $r$th-step normal form $Z^{(r)}$ is quantified by the size of the series function
\begin{equation}
\label{eq:rem}
R^{(r)}=\lambda^{r+1} H^{(r)}_{r+1} + \lambda^{r+2} H^{(r)}_{r+2}+\ldots\ ,
\end{equation}
called the remainder. Estimating the size of the remainder allows to estimate the precision of the normal form analytical approximations to the dynamics (see \citet{giorgilli2002} or \citet{efthyLaPlata} for an introduction to the method of canonical normalization via Lie series). 
According to the method of Lie series, the sequence of canonical transformations \eqref{eq:translie} is determined via the definition of a sequence of Lie generating functions $\chi_1,\chi_2,\ldots$. Namely, after $r$ steps, the transformations are given by:
\begin{eqnarray}
\label{eq:lietra}
\boldsymbol{\varphi}&\equiv\boldsymbol{\varphi}^{(0)}
=&\exp(L_{\chi_{r}})\exp(L_{\chi_{r-1}})\ldots\exp(L_{\chi_1}) \boldsymbol{\varphi}^{(r)} \\ \boldsymbol{J}&\equiv\boldsymbol{J}^{(0)}
=&\exp(L_{\chi_{r}})\exp(L_{\chi_{r-1}})\ldots\exp(L_{\chi_1}) \boldsymbol{J}^{(r)} \nonumber ~,
\end{eqnarray}
where $L_{\chi}\equiv \{\cdot,\chi\}$ is the Poisson bracket operator, and $\exp(L_\chi)=\sum_{k=0}^\infty (1/k!)L_\chi^k$. In practice, we truncate the latter sum (as well as all resulting expressions) at a maximum order $n_{max}$ in the book-keeping parameter $\lambda$. One has to fix $n_{max}$ to a value $n_{max}\geq n_{norm}$, where $n_{norm}$ is the maximum normalization order, i.e., the maximum value of $r$ in the algorithm. 

The task now is to determine the form of the functions $\chi_r$, $r=1,2,\ldots n$.
Assume $r-1$ steps were accomplished. Then, the Hamiltonian is `in normal form' up to the order $r-1$, i.e., $H^{(r-1)} = Z_0+\lambda Z_1+ \ldots + \lambda^{r-1}Z_{r-1}+ \lambda^r H^{(r-1)}_r+\ldots$.
The term $H^{(r-1)}_r$ can now be decomposed as $H^{(r-1)}_r=Z_r+h^{(r-1)}_r$, where $h^{(r-1)}_r$ are the terms which we wish to eliminate from the Hamiltonian, in order to bring the latter in normal form up to order $r$.
Then, the generating function $\chi_r$ accomplishing this task is given by the solution of the {\em homological equation}:
\begin{equation}
\label{eq:homo}
\left\lbrace Z_0, \chi_r \right\rbrace + \lambda^r h^{(r-1)}_r = 0~~.
\end{equation}
The homological equation is possible to solve provided that the function $h^{(r-1)}_r$ is chosen so as to belong to the range of the operator $L_{Z_0}$. Under this condition, the function $h^{(r-1)}_r$ contains a sum of trigonometric monomials
\begin{equation}
h^{(r-1)}_r = \sum_{\boldsymbol{s},\boldsymbol{k}}
a_{\boldsymbol{s},\boldsymbol{k}} J_\rho^{s_1}J_\phi^{s_2}J_z^{s_3}
\exp(k_1\varphi_\rho+k_2\varphi+k_3\varphi_z+k_4\varphi_E+
k_5\varphi_M+k_6\varphi_{M_a}+k_7\varphi_{M_p}+k_8\varphi_{M_s})~,
\end{equation}
where i) $\boldsymbol{s}\equiv(s_1,s_2,s_3)$,  $\boldsymbol{k}\equiv(k_1,k_2,k_3,k_4,k_5,k_6,k_7,k_8)$, and ii) for simplicity in the notation, we drop superscripts indicating the normalization order from all the canonical variables $(\boldsymbol{\varphi},\boldsymbol{J})$.
The solution of the homological equation (\ref{eq:homo}) reads:
\begin{equation}
\label{eq:solhomo}
\chi_r = \sum_{\boldsymbol{s},\boldsymbol{k}} \frac{a_{\boldsymbol{s},\boldsymbol{k}} J_\rho^{s_1}J_\varphi^{s_2}J_z^{s_3}
\exp(i(k_1\varphi_\rho+k_2\varphi+k_3\varphi_z+k_4\varphi_E+
k_5\varphi_M+k_6\varphi_{M_a}+k_7\varphi_{M_p}+k_8\varphi_{M_s}))}{i(k_1\kappa+k_3\kappa_z+k_4\Omega_E+
k_5\Omega_M+k_6\Omega_{M_a}+k_7\Omega_{M_p}+k_8\Omega_{M_s})}~~.
\end{equation}
Finally, after specifying $\chi_r$, we compute the new transformed Hamiltonian via
\begin{equation}
H^{(r)}=\exp(L_{\chi_r})H^{(r-1)}~~.
\end{equation}
This resumes one full step of the canonical normalization algorithm.

\end{document}